\documentclass[twocolumn,showpacs,preprintnumbers,amsmath,amssymb,prb,floatfix]{revtex4}

\usepackage{psfig}
\usepackage{natbib}
\usepackage{bm}

\newcommand{\ket}[1]{\left. \vrule\; #1 \right\rangle}
\newcommand{\braket}[3]{\left\langle #1\;\vrule\; #2 \;\vrule\; #3 \right\rangle}
\newcommand{\scalprod}[2]{\left\langle #1\;\vrule\; #2 \right\rangle}
\newcommand{\vect}[1]{\mathbf{#1}}
\newcommand{\vecti}[1]{\bm{#1}}
\newcommand{\uvect}[1]{\hat{\mathbf{#1}}}
\newcommand{\kk}{{\vect{k}}}
\newcommand{\Tr}{\operatorname{Tr}}

\begin{document}

\bibliographystyle{revtex}


\title{Donor states in modulation-doped Si/SiGe heterostructures}
\author{A. Blom$^1$, M. A. Odnoblyudov$^{1,2}$, I. N. Yassievich$^{1,2}$, and K.-A. Chao$^1$}
\affiliation{$^1$Division of Solid State Theory, Department of Physics, Lund University, S-223 62 Lund, Sweden \\
$^2$A. F. Ioffe Physico-Technical Institute, Russian Academy of Science, 194021 St. Petersburg, Russia}

\begin{abstract}
We present a unified approach for calculating the properties of shallow donors inside or outside heterostructure
quantum wells. The method allows us to obtain not only the binding energies of all localized states of any symmetry,
but also the energy width of the resonant states which may appear when a localized state becomes degenerate with the
continuous quantum well subbands. The approach is non-variational, and we are therefore also able to evaluate the wave
functions. This is used to calculate the optical absorption spectrum, which is strongly non-isotropic due to the
selection rules. The results obtained from calculations for Si/Si$_{1-x}$Ge$_x$ quantum wells allow us to present the
general behavior of the impurity states, as the donor position is varied from the center of the well to deep inside the
barrier. The influence on the donor ground state from both the central-cell effect and the strain arising from the
lattice mismatch is carefully considered.
\end{abstract}

\pacs{73.20.Hb, 73.21.Fg, 78.67.De}


\maketitle


\section{Introduction}\label{sec:intro}

Impurity states in heterostructures have been the subject of detailed investigations during the last two decades.
Traditionally, impurities are considered to be purely detrimental, by increasing the scattering rates. On the other
hand, doping is essential to supply enough free carriers into the system. The strive has therefore been to remove the
doping region from the active region by using modulation doping.

Recently, however, impurities have been placed in the active region of Si/SiGe quantum well structures, to exploit
their presence and properties for novel optical devices in the far-infrared or terahertz (THz)
region.\cite{blomapl,althukov,cambridge} Si and Ge are non-polar materials with low intrinsic absorption at THz
frequencies. Taking into account also the possibilities for integration with existing device technology,\cite{SiReview}
these systems appear very attractive for optical applications in the THz-region, which recently has received a lot of
attention in a variety of fields.\cite{THzReview} A detailed knowledge of the properties of impurity states in Si/SiGe
heterostructures is therefore essential.

Much effort has been spent on calculating the impurity energy states in a quantum well (QW). The techniques, all within
the effective-mass approximation, range from the commonly employed variational
techniques~\cite{bastard,greene1,greene2,imagecharges,betancur,jayakumar,masselink,pasquarello} to direct integration
of the Schr{\"o}dinger equation~\cite{stopa} and basis expansions.\cite{Loehr,aleshkin} The systems considered have, to
the best of our knowledge, exclusively been III--V materials such as GaAs/Al$_{x}$Ga$_{1-x}$As quantum wells, for both
donors~\cite{bastard,greene1,greene2,imagecharges,stopa,betancur,jayakumar} and
acceptors,\cite{pasquarello,Loehr,masselink,aleshkin} and the results are found to agree reasonably well with
experimentally obtained values.\cite{exp1,exp2,exp3,exp4,exp5} Particular aspects such as image
charges\cite{imagecharges} and the effect of screening of the impurity potential by the free carriers in the
well\cite{BrumBastardscreening} have also been investigated.

To allow for a more detailed comparison of the calculated ground state binding energy with experimental results, one
must also take into account the central-cell effect, or the chemical shift. Its physical origin is the absence of
screening by the valence electrons at distances smaller than the outer electronic shells, in which case effective-mass
theory breaks down. In principle, therefore, a proper treatment of chemical shifts is not possible within the
traditional effective-mass theory.\cite{pantelides} Nevertheless, several authors have considered the central-cell
effect in the bulk by introducing various short-range pseudo-potentials, which are adjusted to give agreement with the
experimental binding energies.\cite{ccbulk,ccbulk2,bulk1} This approach has also been attempted in the quantum well
case,\cite{masselink} whereas Mueller {\em et~al.} have chosen to relate the QW central-cell shift to the envelope
function amplitude.\cite{ccqw}

The donor ground state naturally appears below the lowest QW quantization level. As we will see, there are however
Rydberg-like series of impurity states attached to each QW subband, as well as to the three-dimensional continuum. The
situation is thus very similar to what has been observed in strong magnetic fields, where impurity states are attached
to each Landau level.\cite{Wagner}

It has been noted by several authors that in narrow quantum wells, the lowest anti-symmetric impurity state, which is
bound to the second QW subband, may become resonant with the continuum of the first subband.\cite{imagecharges,greene2}
Resonant states of this so-called Fano type, i.e.\ a discrete (and hence localized) state degenerate with a continuum,
are well-known from atomic physics~\cite{Fano} and bulk
semiconductors.\cite{bulk1,bulk2,bulk3,bulk4,Capasso,Capasso2,moprb} If symmetry allows it, the localized state can
couple, or hybridize, with the continuous states. The resonant state is then characterized by an energy width, which is
immediately related to its life-time.\cite{Fano} It is of crucial importance to determine this width in order to
consider the effects of the coupled impurity states on the optical and electrical properties of modulation-doped
quantum wells.

When the impurity is placed outside the well, in the barrier, similar but entirely different resonant states may also
be formed from the usual shallow donor levels, and their widths can be evaluated by the resonant coupling
method.\cite{blomprb} Attempts have been made to apply an essentially equivalent method for the case of donors inside
the well.\cite{jayakumar,Priester,Yen} It has however been shown that the approximations made in this approach are
rather severe, and the widths do not agree with more exact results obtained using the same method as will be presented
in this paper.\cite{blompss} Only when the coupling between the impurity state and the quantum well is weak, can such
perturbation methods be expected to yield accurate results.

In this work we shall present a non-variational approach for calculating the energy levels of shallow donors, in which
we expand the total Hamiltonian in a complete basis. This turns the Schr{\"o}dinger equation into a matrix eigenvalue
problem, which is diagonalized to yield all localized, hybridized and continuous states. The approach has several
benefits over the variational method. First of all, no assumptions are made regarding the form of the impurity wave
functions, but instead we obtain the correct envelope function, which then can be employed for calculating e.g.\
optical matrix elements. Second, the entire energy spectrum is treated simultaneously, and hence we also obtain a
description of the perturbed continuum and can observe the formation of the hybridized resonant states, and calculate
their energy widths. Additionally, the difference in the effective mass in the well and barriers is easily included, as
well as an electric field, and we are able to consider (but not limited to) anisotropic materials such as Si and Ge. It
is furthermore possible to place the impurities in the barrier instead of in the well, and we thus have a unified
approach for treating impurities in modulation-doped heterostructures.

We will specifically consider Si/Si$_{1-x}$Ge$_x$ quantum wells, which are strained due to the lattice constant
mismatch. In result, the sixfold degeneracy of the conduction band bottom, and the donor states, is partially lifted.
Moveover, the impurity ground state in bulk Si is known to be shifted and split by the central-cell
effect.\cite{KohnLuttinger,BiPiBook} We shall therefore also present an approach for simultaneously taking the strain
and the central-cell effect into account.

One of the advantageous properties of a coupled resonant state is the possibility of populating it by electrically
pumping the carriers, from the bottom of the subband, until they reach the resonance
energy.\cite{moprb,resscatt,nanotech} They may then be captured into the resonant state, and possibly make an optical
transition to the ground state. If an inverted population could be arranged between the resonant state and the ground
state, one could realize a laser based on this process. A particularly appealing point of such a device is that since
the impurity states are attached to the QW levels, it is possible to control the intra-impurity transition energy by
varying the QW parameters. Experimental evidence of optical transitions involving coupled resonant states in quantum
wells exists from both Raman scattering~\cite{Perry} and absorption spectroscopy~\cite{Helm} measurements. We therefore
also calculate the impurity absorption spectrum, for arbitrary polarization.

The remainder of our paper is outlined as follows. In Section~\ref{sec:model} we present the non-variational method for
solving the Schr{\"o}dinger equation of a shallow donor in a QW. Once the matrix eigenvalue problem is solved, the
characterization of the eigenstates becomes an important issue, and this is discussed in
Section~\ref{sec:classification}. The influence of strain and the central-cell effect is considered in
Section~\ref{sec:cc}, and in Section~\ref{sec:abs} we calculate the quantum well optical absorption spectrum in the
presence of coupled resonant states. The numerical results of the calculations for Si/Si$_{1-x}$Ge$_x$ quantum wells
are presented and discussed in Section~\ref{sec:results}, followed by a summary in Section~\ref{sec:summary}.


\section{The basis expansion method} \label{sec:model}

We shall consider the problem of a shallow donor, described by the impurity potential $V_c(\vect{r})$, in a quantum
well. Possibly an electric field $\mathcal{E}$ along the QW growth direction $z$ is also present. This field may be
externally applied or built-in due to charge redistribution in the structure. In the effective-mass approximation, the
total Hamiltonian can be written as
\begin{equation}\label{eq:totalH}
\hat{H} = \hat{H}_\mathrm{QW} + V_c(\vect{r}) + e\mathcal{E}z,
\end{equation}
with the quantum well Hamiltonian~\cite{BastardBrum}
\begin{equation} \label{eq:HQW}
\hat{H}_\mathrm{QW} = - \frac{\hbar^2}{2}\left( \frac{\partial}{\partial z}\frac{1}{m_\bot (z)}
\frac{\partial}{\partial z} + \frac{1}{m_{||}(z)} \nabla_{\textrm{2D}}^2 \right) + V(z).
\end{equation}
From a principal point of view, our method allows for the quantum well potential profile $V(z)$ to vary arbitrarily
with the position $z$. We will however, for simplicity, choose $V(z)$ to be a constant band offset $U$ outside the well
(which has width $a$) and zero inside it. The effect of strain due to the lattice mismatch between the well and barrier
regions is assumed to be incorporated in the band offset, except for the splitting of the donor ground state. This will
be considered in detail in Section~\ref{sec:cc}, along with the central-cell potential which is not included in the
effective-mass Hamiltonian~\eqref{eq:totalH}.

We wish to apply our method to anisotropic materials such as SiGe alloys. We have therefore split the kinetic energy
operator into two terms, and let the effective mass depend on the direction, with $m_\bot$ and $m_{||}$ denoting the
respective masses perpendicular and parallel to the two-dimensional (2D) QW plane. Further note, that both the parallel
and perpendicular masses are taken to depend on the coordinate $z$, due to the presence of the heteroboundaries. In
result, the Schr{\"o}dinger equation with the Hamiltonian $\hat{H}_\mathrm{QW}$ will not be separable.

The dependence of the parallel mass $m_{||}$ on $z$ is usually ignored in heterostructure calculations, since the
electrons involved in in-plane transport are strongly confined to the QW channel. In the present case, this
simplification is not appropriate since -- especially for narrow wells -- a large portion of the impurity wave function
may appear in the barriers, and the binding energy depends critically on the exponentially decaying part of the wave
function. The discontinuity of the parallel mass can alternatively be taken into account by introducing an effective
well depth, which depends on the in-plane momentum,\cite{Davies} but this is not convenient for our purposes. By
diagonalizing $\hat{H}_\mathrm{QW}$ within our basis expansion method (see below), we find that a difference in the
parallel masses leads to non-parabolicity of the QW subbands, but does not mix the different QW levels. (This is not to
be confused with the non-parabolicity that arises from coupling to the valence band; c.f.\ the discussion after
Eq.~\eqref{eq:H0energy}.)

The impurity potential is taken as the Coulomb potential
\begin{equation}
V_c(\vecti{\rho},z) = -\frac{e^2}{\epsilon(z)\sqrt{\rho^2+(z-z_0)^2}},
\end{equation}
(in CGS units) as is appropriate for a shallow donor located at $(0,0,z_0)$. Here $\vecti{\rho}$ is the radial vector
in the QW plane. It has been demonstrated that the discontinuity in the dielectric constant $\epsilon$ is of importance
in GaAs/Al$_{x}$Ga$_{1-x}$As systems.\cite{imagecharges} To take this into account properly, one could introduce image
charges, but to evaluate the matrix elements with the resulting effective impurity potential (c.f.\ Eqs.~(5) and (6) in
Ref.~\onlinecite{pasquarello}) would lead to rather extensive calculations in our present approach. Since the
difference in the dielectric constant is small between Si and Si/Si$_{1-x}$Ge$_{x}$ for small $x$, we will assume that
$\epsilon$ is a constant, independent of $z$.

Our method for solving the Schr{\"o}dinger equation $\hat{H} \Psi(\vect{r}) = E \Psi(\vect{r})$ with the total
Hamiltonian~\eqref{eq:totalH}, is based on expanding the total wave function $\Psi(\vect{r})$ in a complete basis, and
diagonalizing the Hamiltonian in this basis. As the basis it is natural to use the quantum well eigenstates
\begin{equation}
\ket{q\kk} = \frac{e^{i\kk\cdot\vecti{\rho}}}{2\pi}\ \varphi_q(z), \quad \scalprod{q'\kk'}{q\kk} =
\delta(\kk-\kk')\delta_{q,q'},
\end{equation}
normalized as indicated, where $q$ enumerates the QW levels and $\kk$ is the wave vector for the in-plane motion. These
states diagonalize
\begin{equation} \label{eq:H0}
\hat{H}_0 = \left[ -\frac{\hbar^2}{2} \frac{\partial}{\partial z}\frac{1}{m_\bot (z)} \frac{\partial}{\partial z}
+V(z)\right] - \frac{\hbar^2}{2m_{||}^w} \nabla_{\textrm{2D}}^2,
\end{equation}
where $m_{||}^w$ is the effective mass inside the quantum well for the direction parallel to the QW plane.

The Hamiltonian $\hat{H}_0$ differs from $\hat{H}_\mathrm{QW}$ in that the parallel effective mass $m_{||}$ does not
depend on the $z$-coordinate. Hence, in contrast to the case with $\hat{H}_\mathrm{QW}$, the Schr{\"o}dinger equation
with the Hamiltonian $\hat{H}_0$ is separable. To be complete, the basis must include both bound and unbound states,
and we therefore enclose the system in a box of width $\mathcal{L}_z$ in the $z$ direction. The box is chosen large
enough that it has negligible influence on the results. Nevertheless, the basis is complete for any size of the box.
The additional boundary condition, that $\varphi_q(z)$ should vanish outside the box, has to be taken into account also
for the bound states to make them properly orthogonal to the unbound states.

The wave functions $\varphi_q(z)$ can be found by standard methods,\cite{Davies} and the energy eigenvalues are given
by
\begin{equation} \label{eq:H0energy}
\hat{H}_0 \ket{q\kk} = E_{qk} \ket{q\kk}, \quad E_{qk} = E_q + E_k,
\end{equation}
where $E_q$ are the energies of the QW levels. In the simplest case the QW subbands are parabolic and $E_k = \hbar^2
k^2/2m_{||}^w   $. In principle one can however introduce here the realistic band dispersion, although it is necessary
that the energy function $E_k$ depends only on the magnitude $k$ of $\kk$. We stress that the discontinuity in the
parallel effective mass $m_{||}$ (see above) is a separate issue; any non-parabolicity in the dispersion of the basis
states is due to coupling to the valence bands and/or higher conduction bands.

Expanding in the complete orthonormal set $\ket{q\kk}$, the total wave function can now be written as
\begin{equation} \label{eq:Psi}
\Psi(\vect{r}) = \sum_{q} \int d\kk\ C_q(\kk) \ket{q\kk}.
\end{equation}
An important benefit of the chosen basis is that $\Psi(\vect{r})$ fulfills the QW boundary conditions by construction.

The spherical symmetry of the Coulomb potential is broken by the presence of the quantum well, and instead the total
system adopts a cylindrical symmetry around the QW growth axis $z$. The total angular momentum $\hat{\vect{L}}^2$ is
hence not conserved, but its projection $\hat{L}_z$ on the growth axis is. The eigenfunctions of $\hat{L}_z$ are
$e^{-im\theta}$ with $m=0,\pm 1,\pm 2,\ldots$, and they form a complete set. The same holds in Fourier space, and hence
we can write
\begin{equation} \label{eq:ckkkk}
C_q(\kk) = C_q(k,\theta_k) = \sum_{m=-\infty}^{\infty} i^{-m} e^{-im\theta_k} C_{qm}(k).
\end{equation}
By inserting this into the expansion Eq.~\eqref{eq:Psi} and performing the angular integral, the eigenstates of the
Hamiltonian $\hat{H}$ given by Eq.~\eqref{eq:totalH} can be expressed as
\begin{equation} \label{eq:expansion}
\psi_m(\rho,\phi,z) = e^{-im\phi} \sum_{q} \varphi_q(z) \int_0^\infty kdk\ C_{qm}(k)\ \mathrm{J}_m(k\rho)
\end{equation}
where $\mathrm{J}_m$ is the $m$-th order Bessel function.

In principle one can diagonalize the total Hamiltonian $\hat{H}$ using this form of the wave function, but it leads to
several integrals which can not be evaluated analytically. Instead, we insert the expansion \eqref{eq:Psi}, taking into
account the angular separation according to Eq.~\eqref{eq:ckkkk}, into the Schr{\"o}dinger equation
$\hat{H}\Psi(\vect{r})=E\Psi(\vect{r})$. As expected from the cylindrical symmetry, the subspaces belonging to
different projections $m$ are not coupled to each other. We can therefore solve the problem -- i.e.\ determine the
expansion coefficients $C_{qm}(k)$ -- separately in each subspace (for a fixed $m$).

The basis expansion turns the Schr{\"o}dinger equation into a Fredholm integral equation of the second kind. It is
customary to symmetrize such equations by multiplying both sides by $\sqrt{k}$ and defining
\begin{equation}
D_{qm}(k) \equiv \sqrt{k}\ C_{qm}(k)
\end{equation}
the integral equation takes the final form
\begin{widetext}
\begin{equation} \label{eq:SEintegral}
\sum_{q'} \int_0^\infty dk'\ h_{qq'}^m(k,k')\ D_{q'm}(k') = (E-E_{qk})\ D_{qm}(k) - \sum_{q'} \Big[
 \mathcal{M}_{qq'}(k) + \mathcal{E}_{qq'} \Big] D_{q'm}(k).
\end{equation}
where the Hermitian kernel
\begin{equation} \label{eq:kernel}
h_{qq'}^m(k,k') = \frac{-e^2\sqrt{kk'}}{2\pi\epsilon} \int_{-\infty}^\infty dz\ \varphi_{q}^*(z)\ \varphi_{q'}(z)
\int_0^{2\pi} d\theta\ \frac{e^{-im\theta}\ e^{-|z-z_0|\alpha_{kk'}(\theta)}}{\alpha_{kk'}(\theta)}
\end{equation}
\end{widetext}
was calculated using the 2D Fourier transform of the Coulomb potential. $\theta$ is the angle between $\kk$ and $\kk'$,
and $\alpha_{kk'}(\theta) = \sqrt{k^2+{k'}^2-2kk'\cos\theta}$. The quantity
\begin{equation} \label{eq:Mqq}
\mathcal{M}_{qq'}(k) = \frac{\hbar^2k^2}{2m_0 m_{||}^w} \int_{-\infty}^\infty dz\ \varphi_q^*(z) \varphi_{q'}(z)
\left(\frac{m_{||}^w}{m_{||}(z)}-1\right)
\end{equation}
takes care of the discontinuity of the parallel mass $m_{||}(z)$, and the electric field $\mathcal{E}$ enters through
\begin{equation}
\mathcal{E}_{qq'} = e\mathcal{E}\int_{-\infty}^\infty dz\ \varphi_q^*(z)\ z\ \varphi_{q'}(z).
\end{equation}

To solve the integral equation we discretize the $k$-axis and approximate the integral over $k$ by a discrete sum. This
turns the integral equation \eqref{eq:SEintegral} into a matrix eigenvalue problem
$\mathcal{H}_{qq'}^{kk'}\mathcal{D}=E\mathcal{D}$, where $\mathcal{D}$ is a column vector of the (renormalized; see
below) coefficients $D_{qm}(k)$. In the simplest (but most convenient) case we choose an equal stepsize $\Delta k$ for
the discretization, and the matrix elements are then given by
\begin{equation} \label{eq:matrixelement}
\begin{split}
\mathcal{H}_{qq'}^{kk'} = & \; \Big( E_{qk} \delta_{q,q'} + \mathcal{M}_{qq'}(k) + \mathcal{E}_{qq'} \Big)
\delta_{k,k'}
\\[+2mm]&+h_{qq'}^m(k,k') \Delta k.
\end{split}
\end{equation}

Due to the long range nature of the Coulomb potential, a singularity appears in the kernel Eq.~\eqref{eq:kernel} for
scattering in the forward direction. We do not explicitly consider screening in this work; instead the divergence is
handled by averaging over small scattering angles.\cite{Loehr} In the case $k=k'=0$ the entire kernel vanishes exactly,
but for other values of $k=k'$ we obtain
\begin{widetext}
\begin{equation}
h_{qq'}^m(k,k) \approx \frac{-e^2 \sqrt{kk}}{2\pi\epsilon} \int_{-\infty}^\infty dz\ \varphi_{q}^*(z)\ \varphi_{q'}(z)
\int_{\Delta k/2k}^{2\pi-\Delta k/2k} d\theta\ \frac{e^{-im\theta}\
e^{-|z-z_0|\alpha_{kk}(\theta)}}{\alpha_{kk}(\theta)} + \frac{-e^2}{\epsilon\sqrt{\pi}}\ \delta_{q,q'}. \qquad (k \neq
0)
\end{equation}

A word on the normalization of the wave functions is appropriate. The matrix eigenvalue problem is solved by standard
numerical methods, which return the eigenvectors $\mathcal{D}^{(i)}$ and eigenvalues $E^{(i)}$ for each eigenstate $i$.
Typically such eigenvectors are normalized by $\sum_{qk}|\mathcal{D}_{qm}^{(i)}(k)|^2=1$. Using the real-space density
function
\begin{equation} \label{eq:density_def}
\begin{split}
P^{(i)}(\vect{r}) &\equiv \int d\vect{r}'\ [\psi_m^{(i)}(\vect{r}')]^*\ \delta(\vect{r}-\vect{r}')\ \psi_m^{(i)}
(\vect{r}) \\[+1mm]%
&= \int_0^\infty dk\ \sqrt{k}\ \mathrm{J}_m(k\rho) \int_0^\infty dk'\ \sqrt{k'}\ \mathrm{J}_m(k'\rho)\ \sum_{qq'}
[D_{qm}^{(i)}(k)]^* D_{q'm}^{(i)}(k')\ \varphi_q^*(z)\ \varphi_{q'}(z),
\end{split}
\end{equation}
\end{widetext}
the normalization condition of the corresponding wave function $\psi_m^{(i)}(\vect{r})$ becomes (after discretization
$\int dk \to \sum_k\Delta k$),
\begin{equation}
1 = \int P^{(i)}(\vect{r}) d\vect{r} = 2\pi\Delta k \sum_{qk} \left| D_{qm}^{(i)}(k) \right|^2.
\end{equation}
Thus the wave function $\psi_m^{(i)}$ will be properly normalized if the physical coefficients $D_{qm}^{(i)}(k)$ are
taken as
\begin{equation} \label{eq:normDD}
D_{qm}^{(i)}(k) = \frac{\mathcal{D}_{qm}^{(i)}(k)}{\sqrt{2\pi\Delta k}}.
\end{equation}

For the particular basis and QW potential profile we have chosen, most integrals which appear in the expressions above
can be evaluated analytically, also when an electric field is included, leaving only a numerical integral over the
angle $\theta$. The expressions are however much too lengthy to include here.


\section{Characterization of the eigenstates} \label{sec:classification}

The diagonalization of the matrix problem provides the energies of all eigenstates of the
Hamiltonian~\eqref{eq:totalH}, and using Eq.~ \eqref{eq:expansion} the corresponding eigenvectors allow us to evaluate
the wave functions and hence matrix elements such as optical dipole-interaction strengths. It is however not
immediately obvious how to identify the individual eigenstates as localized impurity states, QW band states or
hybridized states, and we shall now address this question.

We will adopt the notation of Ref.~\onlinecite{imagecharges} and denote the case $m=0$ by $\Sigma$ and all other values
by $\Pi$. If the impurity is placed in the center of the well, and there is no electric field present, the system is
invariant under reflections $z\ \to\ -z$, and each eigenstate will also possess a quantum label $g$ and $u$ for even
and odd parity, respectively. If this symmetry is broken, the only good quantum labels will be $m$ and the energy $E$.
It is however almost always possible to trace the eigenstates back to the symmetric situation, and we will therefore
use eigenstate labels such as $\Sigma_g$ even when asymmetry is present.

Our interest is particularly focused on the lowest anti-symmetric impurity state, and we distinguish it from the other
states of the same symmetry by referring to it as the $\Sigma_u^*$ state. This state has been shown to be attached to
the second QW subband, and for narrow well widths it becomes resonant with the continuum of the first (lowest)
subband.\cite{imagecharges,greene2} It was further demonstrated that the $m=\pm 1$ or $\Pi_u$ states do not become
resonant, but instead remain attached to the lowest subband. We will therefore from now on specialize to the case $m=0$
and often omit the label $m$.

In the absence of an electric field, we can always identify the states corresponding to the original QW levels (the
subband bottoms), from the fact that these are the only eigenstates with non-zero contribution from the $k=0$ basis
states. To facilitate the further classification, we define for each eigenstate $i$, with energy $E_i$, a quantity
\begin{equation}
d_q(E_i) = 2\pi \int_0^\infty dk\ |D^{(i)}_q(k)|^2,
\end{equation}
which measures the relative contribution to that eigenstate from the $q$-th basis state $\varphi_q(z)$; hence $\sum_q
d_q=1$ for all states. The lowest QW level is taken as $q=0$.

Studying the values of $d_q(E_i)$ for a particular eigenstate $i$ now allows us to distinguish between three types of
states:
\begin{itemize}
  \item {\em Localized impurity states} only have contributions from higher QW levels, i.e.\ non-zero $d_q$ only for levels
  $q$ with energy $E_q>E_i$.
  \item {\em Continuous QW subband states} only have contributions from lower levels, $E_q<E_i$. (This is valid also in the
  case of a discontinuous parallel effective mass.)
  \item {\em Hybridized (coupled) resonant states} have contributions from both higher and lower levels.
\end{itemize}

\begin{figure}[b]
\centerline{\psfig{figure=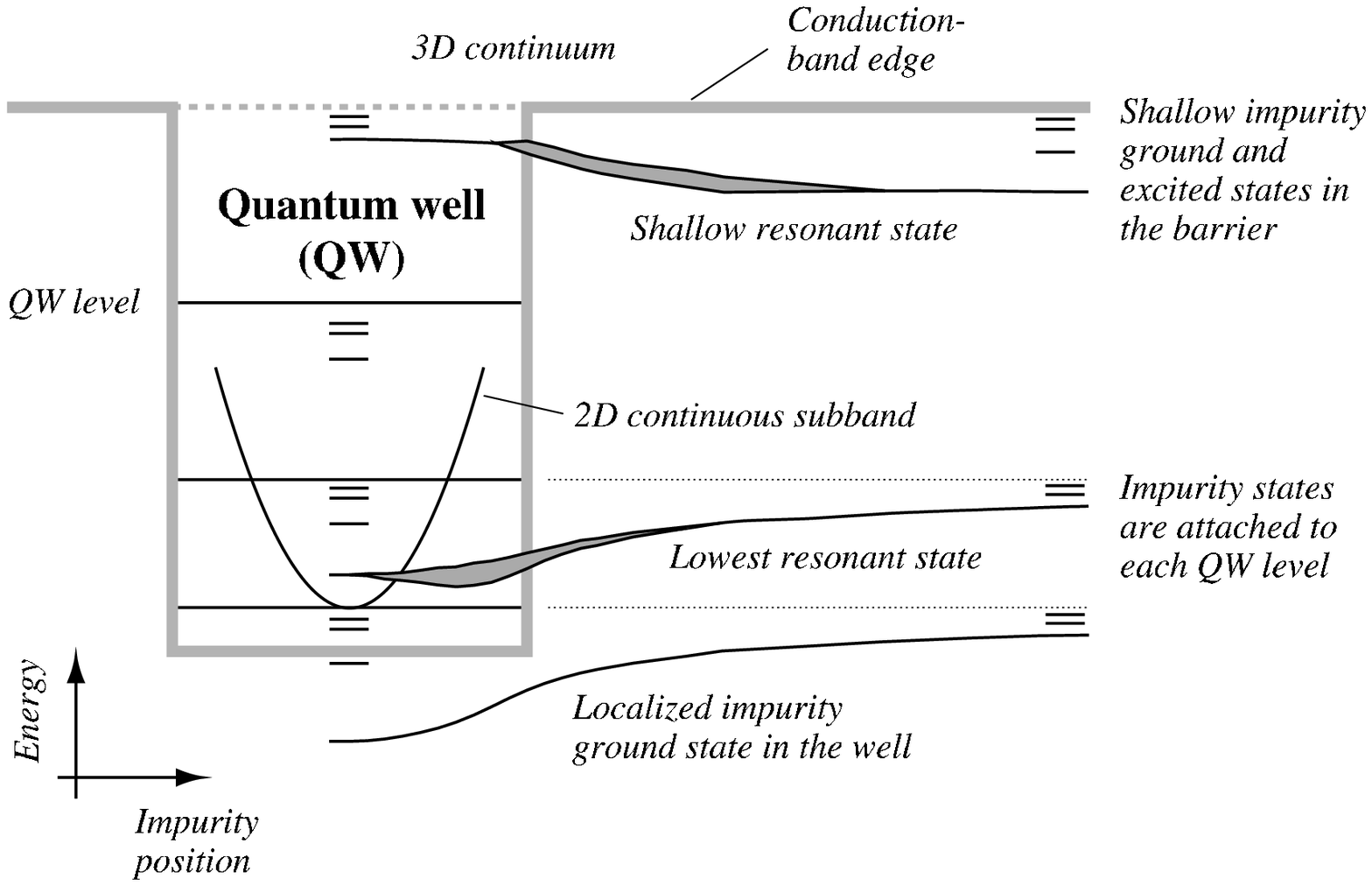,width=\linewidth}}%
\caption{Schematic picture, presenting the general behavior of how the various impurity states evolve as we change the
impurity position, for a donor placed inside or outside a heterostructure quantum well. The shaded areas represent the
energy width of the resonant states.}\label{fig:qw}
\end{figure}

Our method allows us to place the impurity anywhere in the system. Figure~\ref{fig:qw} summarizes the qualitative
behavior of localized and resonant impurity states, as we change the impurity position from the middle of the quantum
well to a remote location in the barrier. Some properties displayed schematically in this figure will be considered in
more quantitative details in Section~\ref{sec:results}, whereas the remainder of this section will focus on the
properties which characterize the various impurity states.

Any state appearing in the energy region below the lowest QW level must be localized. The lowest one will be the
impurity ground state, which is split due to the central-cell effect (c.f.\ Section~\ref{sec:cc}). There are several
$\Sigma_g$ states attached to the lowest subband, and they form what resembles a Rydberg series, with decreasing
binding energies converging towards the lowest subband edge. Actually, such series of localized states appear below
each subband, i.e.\ each QW level has a set of impurity states attached to it (c.f.\ Figure~\ref{fig:qw}). The binding
energy of a localized state is therefore to be understood as the smallest energy required to place an initially
localized electron into the corresponding QW subband. It is noteworthy that these considerations also apply to the
unbound QW levels. Hence if the well is so narrow that there exists no second bound QW level, all the $\Sigma_u$ states
are still well-defined, but attached to the three-dimensional (3D) continuum.

If the quantum well is wide enough, so that the binding energy of the lowest resonant state $\Sigma_u^*$ is larger than
the separation between the first ($q=0$) and second ($q=1$) QW levels, this state appears below the lowest QW level,
and is therefore localized. It is, however, still attached to the $q=1$ level. As the well width is decreased, the
$\Sigma_u^*$ state will therefore eventually appear in the continuum of the lowest QW subband, and we obtain a Fano
resonant state -- a localized state degenerate with a continuum. Similar resonant states can be formed from any excited
impurity level (except those attached to the lowest subband), for suitable well parameters. We shall limit most of the
discussion to the $\Sigma_u^*$ state, but our approach can equally well be applied to any of the resonant states, of
either parity.

\begin{figure}
\centerline{\psfig{figure=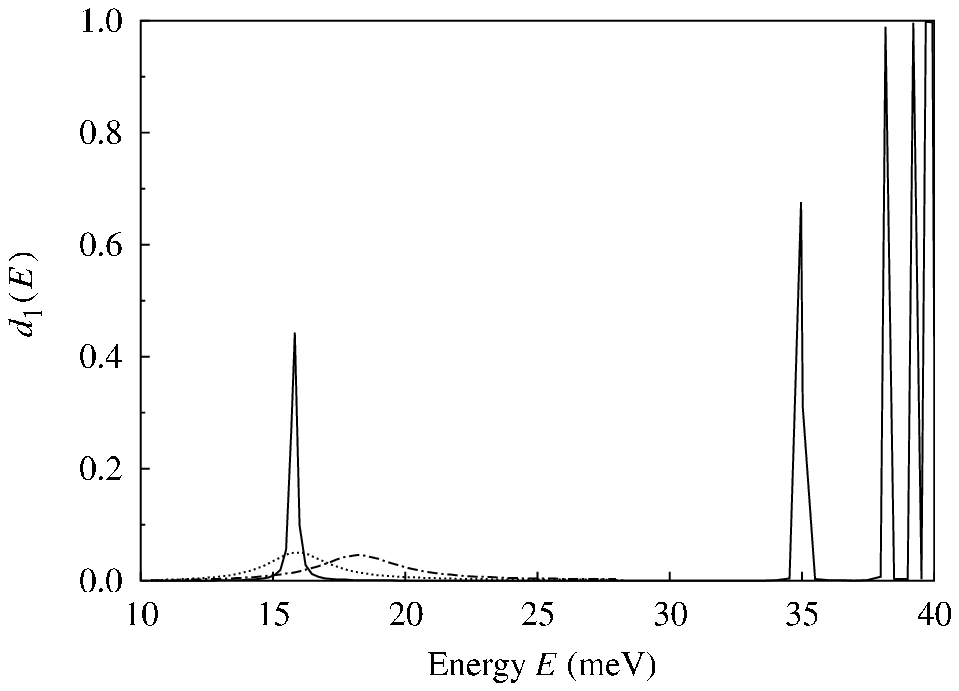,width=\linewidth}}%
\caption{The contribution $d_1(E)$ of the second QW subband to the eigenstates in the energy range from the first QW
level (at 10~meV) to the second one (at 40~meV). The width of the Si/Si$_{0.9}$Ge$_{0.1}$ well is $a$=50~\AA, and the
three curves correspond to different impurity positions $z_0$: $z_0/a$=0.05 (solid), 0.25 (dotted), and 0.45
(dash-dot); $z=0$ is the middle of the well. In addition to the resonances at 15--20~meV, similar ones resembling a
Rydberg series, appear close to the bottom of the second subband (for clarity this is only shown for $z_0/a$=0.05).}
\label{fig:contrib}
\end{figure}

As long as the impurity is placed in the center of the well, no hybridization or coupling can take place between
impurity states which are resonant with subbands of opposite parity. The localized $\Sigma_u^*$ state is attached to
the $q=1$ QW level, and therefore $d_0(E_{\Sigma_u^*})=0$ in the symmetric situation. Moreover, without any
hybridization, $d_q=\delta_{q,0}$ for the continuous states belonging to the lowest QW level. Coupling will however be
present as soon as any asymmetry which breaks the parity conservation is introduced -- such as shifting the impurity
position or applying an electric field. The localized state is then "diluted" throughout a band of actual stationary
states, whose profile is represented by a Lorentzian resonance curve.\cite{Fano} This hybridization mixes a certain
amount of continuum into the $\Sigma_u^*$ state (giving $d_0(E_{\Sigma_u^*})\neq 0$), and the continuous states acquire
a partly localized character ($d_1\neq0$). Thus the degree of hybridization is measured by the coefficients $d_1(E_i)$,
and the resonance profile is exactly given by these coefficients when plotted against the eigenenergies $E_i$ in the
region between the first and second QW levels (c.f.\ Figure~\ref{fig:contrib}). If the electron is initially placed in
the coupled resonant impurity state, it will autoionize with a mean life-time $\tau=\hbar/\Gamma$ determined by the
energy width $\Gamma$ of the resonant state.\cite{Fano} We can calculate this width by fitting a Lorentzian to the
resonance profile.

In the barrier, the binding energies -- which are still measured relative the corresponding QW levels -- decrease
towards zero, although not very rapidly. As the impurity is moved farther away from the well, each impurity state
remains bound to its initial QW level, with vanishing binding energy. Hence, even at very large impurity distances, the
impurity "ground state" appears to be a deep state with an energy defined by the lowest QW level. This is
counter-intuitive; we expect the impurity ground state to be a shallow state bound to the 3D continuum. However, the
electron density of the deep impurity states is actually localized close to the center of the well, and not on the
impurity. When the distance to the well increases, their wave functions become more and more symmetric around the
center of the well, and assumes the shape of the QW basis functions $\varphi_q(z)$. At the same time these wave
functions are extended in the radial direction in the QW plane, due to the smaller binding energy. This behavior can be
understood from the fact that the electronic density is determined by the effective potential in the well, and when the
impurity is far away, the tail of the Coulomb potential is much weaker than the QW confinement potential. At large
impurity distances, the Coulomb potential and the well are almost (but not completely) decoupled, which means that the
deep donor states essentially derive from a single QW level, which can be verified by studying the coefficients $d_q$.
Similar observations have been made by Stopa and DasSarma~\cite{stopa} who also use a non-variational technique to
treat donors in GaAs quantum wells.

As was stated above, there are also QW impurity states attached to the 3D continuum (c.f.\ the upper right part of
Figure~\ref{fig:qw}). When the impurity is in the barrier, these states are the familiar ground and excited shallow
donor states, with the electron density localized at the impurity position. We can now study what happens to an
initially decoupled shallow donor, as it is moved closer to the quantum well. Due to the induced coupling, Rydberg-like
series of impurity states -- localized in the well -- are pushed down from each continuum edge, i.e.\ from each QW
level. The lowered symmetry due to the presence of the quantum well causes mixing between the hydrogenic $n$ and $\ell$
levels (actually only between levels for which $\Delta \ell$ is an even number~\cite{stopa}), and only $m$ remains as a
good quantum label. The original shallow donor state broadens due to the coupling,\cite{blomprb} but remains bound to
the 3D continuum also when the impurity finally is located in the well. If the electron is initially placed in the
shallow donor state outside a quantum well, it will after some time tunnel into the QW. The tunnelling time is
essentially given by the energy width of the resonant shallow state, which can be calculated within our basis expansion
approach, or as in Ref.~\onlinecite{blomprb}.


\section{Strain and central-cell effects}\label{sec:cc}

In order to compare calculated results for the ground state binding energy of donors in Si quantum wells with
experimental values, it is essential to take into account the effects of both strain and the central-cell shift.
Although there would, in principle, be some central-cell splitting also of the excited states, this effect is expected
to be very small.\cite{KohnLuttinger} This is especially true for the odd parity states, since they have a vanishing
envelope function at the impurity position. Thus the energy of the resonant $\Sigma_u^*$ state, which in the bulk limit
corresponds to the $2p_0$ hydrogenic state, will not be affected, and we shall in this section only be concerned with
the impurity ground state.

The conduction band minimum in bulk Si lies at about $\Delta_1=(0,0,0.85)$ in units where the $X$ point is (0,0,1),
with six equivalent valleys. Hence the donor $1s$ ground state would be sixfold degenerate. Experimentally, however,
one instead observes three levels, with binding energies $E_b+\Delta_0$ (the non-degenerate true ground state),
$E_b+\Delta_0-\Delta_E$ (twofold degenerate) and $E_b+\Delta_0-\Delta_T$ (threefold degenerate). Here $E_b$ is the
effective-mass binding energy, which according to Kohn and Luttinger~\cite{KohnLuttinger} is about 29~meV in Si. The
energies $\Delta_0$, $\Delta_E$ and $\Delta_T$ are positive and, in contrast to $E_b$, depend on the particular
impurity species. This additional contribution to the binding energy is known as the central-cell effect or chemical
shift, and is in general determined from comparisons with experimental values. In the case of P donors in Si,
$\Delta_0\approx 16.6$~meV (or 14.3~meV if one uses the effective-mass value $E_b$=31.27~meV from the more elaborate
calculations by Faulkner~\cite{faulkner}), $\Delta_E\approx 13.0$~meV and $\Delta_T\approx 11.7$~meV.\cite{Sidonors}

The chemical shift is usually considered as arising from a strong potential acting only very near the impurity center.
Although, as discussed in the Introduction, a proper treatment is not strictly possible in effective-mass theory, one
could try to incorporate the central-cell effect by employing a short-range pseudo-potential. This, however, fails when
applied to a basis expansion method, since this potential now interacts not with the total wave function (as is the
case in a variational calculation) but with the individual basis states. If one adjusts the pseudo-potential parameters
to produce agreement with experimental binding energies for a certain well width (such as a very wide one, close to
bulk), the parameters will not give any sensible results when the basis is different, say for a narrow well. Instead,
we will here show how to relate the QW central-cell shift to the bulk shifts, purely by symmetry considerations.

In the effective-mass approximation, the wave functions are written as $\Phi(\vect{r})=\phi(\vect{r})u(\vect{r})$,
where the envelope function $\phi$ varies slowly over inter-atomic distances, and the Bloch function $u$ is rapidly
varying and describes the region close to the atomic nuclei. In Si there are six equivalent conduction band minima or
valleys $\ell$, which we index by $\ell\in\{x,\overline{x},y,\overline{y},z,\overline{z}\}$ ($\overline{x}$ is
shorthand for $-x$). Hence one must generally consider linear combinations $u(\vect{r})=\sum_\ell \alpha_\ell
u_\ell(\vect{r})$ of the Bloch functions $u_\ell$ of each valley bottom. If we denote the (unspecified) central-cell
potential by $V$, the non-zero matrix elements between the valley bottom Bloch functions in Si are~\cite{Gastev}
\begin{equation} \label{eq:ccV}
\braket{u_\ell}{V}{u_{\ell'}} =
\begin{cases} V_0, \qquad \ell=\ell', \\%
V_g, \qquad \ell=-\ell', \\%
V_f, \qquad |\ell|\neq|\ell'|.
\end{cases}
\end{equation}
By evaluating the matrix elements $\braket{\Phi_{1s}^i}{V}{\Phi_{1s}^i}$ for the six $1s$ donor states
$\Phi_{1s}^i(\vect{r})$, we can relate the parameters $V_0$, $V_f$ and $V_g$ to the observed impurity binding energies
as follows:
\begin{equation} \label{eq:VfromDelta}
\begin{split}
V_0\ |\phi_{1s}(\vect{r}_0)|^2 &= (3\Delta_T+2\Delta_E-6\Delta_0)/6, \\%
V_g\ |\phi_{1s}(\vect{r}_0)|^2 &= (2\Delta_E-3\Delta_T)/6, \\%
V_f\ |\phi_{1s}(\vect{r}_0)|^2 &= -\Delta_E/6,
\end{split}
\end{equation}
where $\vect{r}_0$ is the impurity position and $\phi_{1s}$ is the $1s$ envelope function. For further details, see
Appendix~\ref{apx:cc}.

We now consider donors placed in a Si quantum well, grown pseudo\-morphically between (unstrained) Si$_{1-x}$Ge$_x$
barriers. This case differs compared to bulk Si in two ways. First, the Si region will be strained due to the lattice
constant mismatch between Si and Ge. Second, the envelope function is different, due to the additional confinement. On
the other hand, it is reasonable to assume that the matrix elements $V_0$, $V_g$ and $V_f$ are unchanged, since the
central-cell potential $V$ is appreciable only in a region much smaller than the well width. Hence, we can still use
the values of these parameters, as obtained from Eq.~\eqref{eq:VfromDelta} using the bulk binding energies (determined
by $\Delta_0$, $\Delta_E$ and $\Delta_T$), but replace $\phi_{1s}$ by the QW envelope function.

By taking into account the effects of strain and the central-cell shifts simultaneously, one finds -- in addition to an
overall shift -- that for (001)-grown wells, the sixfold degenerate donor ground state splits into five energy levels,
whereof one is twofold degenerate.\cite{BiPiBook} The details of this calculation are presented in
Appendix~\ref{apx:largestrain}. The lattice constant of unstrained Si$_{1-x}$Ge$_x$ bulk alloy has been
parametrized~\cite{rieger} from experimental data~\cite{sigelatticeconst} as
\begin{equation*}
a(x) = a(\mathrm{Si}) + 0.0200326 x(1-x) + [a(\mathrm{Ge})-a(\mathrm{Si})]x^2,
\end{equation*}
where $0\leq x \leq 1$ and $a$(Si) and $a$(Ge) are the lattice constants of bulk Si and Ge (in nanometers). Inserting
this into Eqs.~\eqref{eq:latticeconstants} and \eqref{eq:straintensorcomponents} we find that the strain component
parallel to the QW interfaces $e_{||}>0$, and hence the strain parameter $\xi=-\Xi_u e_{||} (2C_{12}/C_{11}+1)/3$
defined in Eq.~\eqref{eq:xi} is negative. Here $\Xi_u$ is the deformation potential and $C_{ij}$ are the components of
the stiffness tensor. Even with only a small content $x$ of Ge, the condition that the strain splitting is large
compared to the central-cell shifts ($\xi \gg \Delta_0$) is fulfilled.\cite{rieger} Therefore, from the results in
Appendix~\ref{apx:largestrain}, the lowest valleys will be $\Delta_\bot$, perpendicular to the QW interfaces, to which
two donor states are attached. Four other donor states are associated with the higher $\Delta_{||}$ valleys, which are
strain split from $\Delta_\bot$ by $3|\xi|$.

In our calculations we assume that the strain splitting of the conduction band is included in the band offset $U$, and
consider only the two lowest donor states, belonging to $\Delta_\bot$. These two states are shifted from the QW
effective-mass value $E_b$ by
\begin{equation} \label{eq:ccqwshift}
\frac{|\psi_\mathrm{QW}(\vect{r}_0)|^2}{|\phi_{1s}(\vect{r}_0)|^2} \left[ - \Delta_0 + \left\{
\begin{matrix} 2\Delta_E/3 \\ \Delta_T \end{matrix} \right\} \right].
\end{equation}
The Bloch functions of the two states are the symmetric and anti-symmetric combinations of the two $\Delta_\bot$
valleys. The quantum well envelope function amplitude $|\psi_\mathrm{QW}(\vect{r}_0)|^2$ is obtained from the basis
expansion coefficients using Eq.~\eqref{eq:density_def}. We will return to the question how to correctly determine the
bulk amplitude $|\phi_{1s}(\vect{r})|^2$ in Section~\ref{sec:results}.

The central-cell shifts are negative, which means that the binding energies are increased. The impurity ground state
now corresponds to the (non-degenerate) $2\Delta_E/3$ state, since $\Delta_E\simeq\Delta_T$ for typical donors (P, As,
Sb) in Si.\cite{Sidonors} It is noteworthy that, since furthermore $\Delta_E\simeq\Delta_0$, the central-cell effect in
the quantum well case can be much smaller than in bulk Si, at least if the ratio of the wave functions is not too
large. In bulk Si, the donor ground state is shifted by $\Delta_0=17$~meV, but in the QW the ground state shift for P
donors is only $\Delta_0-2\Delta_E/3\approx8$~meV, assuming that the envelope function ratio in
Eq.~\eqref{eq:ccqwshift} is of the order unity. On the other hand, this ratio is expected to be above unity for narrow
wells, due to the additional confinement, and so in this case the central-cell effect further increases the already
enhanced ground state binding energy.


\section{Optical absorption} \label{sec:abs}

If the impurity is placed exactly in the middle of the well (and no electric field is present), the selection rules
prohibit the radiative decay from the lowest subband to the impurity ground state by a dipole transition with
polarization parallel to the QW growth direction. However, if some asymmetry is introduced, the transition becomes
allowed from the part of the subband which is hybridized with the resonant anti-symmetric $\Sigma_u^*$ impurity state.

In the dipole approximation, Fermi's Golden Rule gives the probability per unit time of optical absorption, at the
frequency $\omega$, between states $i$ (initial) and $j$ (final) as
\begin{equation} \label{eq:Wabssingle}
W_{ji} = \frac{2\pi}{\hbar} \left(\frac{eA_0}{m c}\right)^2 \left| \braket{j}{\uvect{e}\cdot \hat{\vect{p}} }{i}
\right|^2 \delta(E_j-E_i- \hbar\omega)
\end{equation}
where $A_0$ is the magnitude of the electromagnetic vector potential, $\uvect{e}$ the photon polarization vector,
$\hat{\vect{p}}$ the momentum operator, and $E_i$ the energy of eigenstate $i$.

Assuming that the system contains $N$ independent impurities, we may relate the absorption rate $W_{ji}$ to the
absorption coefficient $\mathcal{A}$ as
\begin{equation}
\mathcal{A}(\hbar\omega) = \frac{N}{\Omega} \frac{2\pi\hbar c}{\kappa \omega |A_0|^2} \sum_j W_{ji},
\end{equation}
where $\Omega$ is the total sample volume and $\kappa$ the refraction index. In a quantum well, a more natural quantity
to consider is the absorption cross-section
\begin{equation} \label{eq:A}
\sigma = \frac{\mathcal{A}\Omega}{N} = \frac{4\pi^2\hbar^2\alpha}{\kappa} \sum_j \left|
 \braket{j}{\frac{\uvect{e}\cdot\hat{\vect{p}}}{m}}{i} \right|^2 \frac{\delta(\hbar\omega_{ji}-\hbar\omega)}{\hbar\omega_{ji}}
\end{equation}
where $\omega_{ji}=(E_j-E_i)/\hbar$ and $\alpha\equiv e^2/\hbar c$ is the fine-structure constant.

The summation in Eq.~\eqref{eq:A} in principle involves, among other things, final states $j$ in all valleys with any
spin polarization. We can however ignore the four $\Delta_{||}$ valleys altogether, since the strain splitting is much
larger than the impurity binding energy. We shall consider absorption only from the ground state, and so the initial
state $i$ is an equal mix of the two $\Delta_\bot$ valleys, as discussed in Section~\ref{sec:cc}, with fixed (but
arbitrary) spin. The dipole operator $\uvect{e}\cdot\hat{\vect{p}}$ does not affect spin, and connects only the parts
of the initial and final states belonging to the same valley. This holds also when the temperature is large enough that
the occupation probabilities of the two states $u_2$ and $u_6$ (c.f.\ Appendix~\ref{apx:largestrain}) are roughly
equal; in this case we must also average over the two states. Thus the sum over $j$ can be restricted to a sum over the
final energy $E$ and the different cylindrical subspaces. It is furthermore enough to evaluate the matrix element
between the envelope functions only, as long as we replace the electron mass $m$ by the effective mass $m^*$ in the
direction parallel to $\hat{\vect{p}}$.\cite{Davies}

Since one would most naturally place the donors inside the QW for optical applications of this kind, we assume that the
relevant effective masses are those of the well, and write
\begin{equation} \label{eq:ep}
\frac{\uvect{e}\cdot \hat{\vect{p}}}{m^*} = \frac{\cos\theta \hat{p}_x}{m_{||}^w} + \frac{\sin\theta
\hat{p}_z}{m_\bot^w},
\end{equation}
where $\theta$ is the angle between the photon $\kk$-vector and the QW growth direction $z$. The light is assumed to be
plane polarized in the plane spanned by the normal to the QW plane and the photon $\kk$-vector. The $x$-axis is defined
as the intersection of this plane and the QW interface planes.

In principle Eq.~\eqref{eq:ep} allows for interference between the two terms, when inserted into Eq.~\eqref{eq:A}. The
interference terms are essentially $\braket{i}{\hat{p_x}}{j}\braket{j}{\hat{p}_z}{i}$ or the complex conjugate thereof,
to be summed over all final states $j$. Here $i$ is the impurity ground state, which belongs to the $m_i=0$ cylindrical
subspace. The sum over final states can be split into two parts, where we first consider all states $j$ in a particular
subspace $m_j$, and then sum over all subspaces. The first factor is $\braket{i}{\hat{p}_x}{j}\propto
\delta_{m_j,m_i\pm 1}$, which follows from the selection rule $\Delta m=\pm 1$ (see below) for the $\hat{p}_x$
operator. On the other hand, $\braket{j}{\hat{p}_z}{i}\propto \delta_{m_j,m_i}$ since $\Delta m=0$ for $\hat{p}_z$.
Thus, the interference terms vanish, and we may write
\begin{equation} \label{eq:absangulardep}
\sigma = \sigma_x \cos^2\theta + \sigma_z \sin^2\theta,
\end{equation}
which defines $\sigma_x$ and $\sigma_z$. As we will see in Section~\ref{sec:results}, the absence of interference is a
contributing factor to producing a symmetric absorption peak, contrary to what is often observed for optical
transitions involving resonant states.\cite{Capasso2,Fano}

Let us first consider $z$-polarization. To eliminate the $\delta$-symbol in \eqref{eq:A} we turn the sum over final
states $j$ into an integral. However, we have to consider that the relevant density of final states in this case is not
the usual 2D density of states. If we double the "size" of our system ($\Delta k \to \Delta k/2$) we only get twice as
many eigenstates, since the discretization in Section~\ref{sec:model} is carried out over the magnitude $k$ of
$\vect{k}$. Hence the density of states in $k$-space is $dn/dk=1/\Delta k$, and in energy space
\begin{equation}
\frac{dn}{dE_k} = \frac{dn}{dk} \frac{dk}{dE_k} = \frac{1}{\hbar\Delta k} \sqrt{\frac{m_{||}^w}{2E_k}}
\end{equation}
where the parabolic dispersion relation $E_k = \hbar^2 k^2/2m_{||}^w$ was again assumed. The presence of $\Delta k$
will properly renormalize the matrix element when this is evaluated from the discretized expansion coefficients (c.f.\
the discussion on normalization in Section~\ref{sec:model}). Now we may perform the integral over energy to remove the
$\delta$-function, and the absorption cross-section can be evaluated at $\omega=\omega_{ji}$ for all eigenstates $j$.
The result is
\begin{equation}
\sigma_z = \frac{2\pi^2\alpha}{\kappa\Delta k}\ \frac{\sqrt{2\hbar m_{||}^w}}{\hbar(m_\bot^w)^2\omega_{ji}^{3/2}}
\big|\braket{j}{\hat{p}_z}{i}\big|^2.
\end{equation}

By using the form Eq.~\eqref{eq:expansion} of the wave function for the states $i$ and $j$ (belonging to the subspaces
$m$ and $m'$, respectively), the remaining matrix element becomes
\begin{widetext}
\begin{equation}
\braket{j}{\hat{p}_z}{i} = \int_0^{2\pi} d\phi\ e^{-i(m-m')\phi} \sum_{qq'} \braket{q'}{\hat{p}_z}{q} \int_0^\infty dk\
D_{qm}^{(i)}(k)\ [D_{q'm'}^{(j)}(k)]^*.
\end{equation}
\end{widetext}
Since the integral yields $2\pi\delta_{m,m'}$, we obtain the selection rule $\Delta m=0$ for this polarization. The
integral over $k$ can be evaluated from the matrix eigenvectors $\mathcal{D}$, taking the normalization relation
Eq.~\eqref{eq:normDD} into account. We can furthermore use the commutator $[\hat{H}_{\mathrm{QW}},z] = -i\hbar
\hat{p}_z/m_{\bot}(z)$ and the fact that the states $q$ and $q'$ are eigenstates of $\hat{H}_{\mathrm{QW}}$, defined in
Eq.~\eqref{eq:HQW}, to rewrite
\begin{equation}
\braket{q'}{\hat{p}_z}{q} = \frac{i}{\hbar}\big(E_{q'}-E_{q}\big) \int_{-\infty}^\infty \varphi_{q'}^*(z)\ z\
m_{\bot}(z)\ \varphi_q(z)\ dz.
\end{equation}

Next we consider $x$-polarization, for which the selection rule is $\Delta m=\pm 1$, as will be demonstrated shortly.
We will focus on the energy region between the first and second QW subbands, where the lowest resonant state
$\Sigma_u^*$ appears. Since there are no resonant states with $m=\pm 1$ in this region, and since the ground state
belongs to the $m=0$ subspace, we can ignore mixing of the QW subbands and represent the final states as normalized
plane waves belonging to the first subband $q=0$:
\begin{equation} \label{eq:jplanewave}
\ket{j} = \frac{\exp(i\kk^{(j)}\cdot\vecti{\rho})}{\mathcal{L}}\ \varphi_{q=0}(z),
\end{equation}
where $\mathcal{L}$ is the normalization length.

To derive the selection rule, one may rewrite the momentum operator $\hat{p}_x$ in cylindrical coordinates and act on
the cylindrical expansion of a plane wave $e^{i\kk\cdot\vecti{\rho}} = \sum_m i^m e^{-im\vartheta}
\mathrm{J}_m(k\rho)$, where $\vartheta$ is the angle between the vectors $\kk$ and $\vecti{\rho}$. One then obtains
\begin{equation} \label{eq:pxmatrixelement}
\braket{j}{\hat{p}_x}{i} = \frac{\pi\hbar k_x^{(j)}}{\mathcal{L}}\ C^{(i)}_{q=0}(k^{(j)}) \sum_{m=-\infty}^\infty
\delta_{m,\pm 1}
\end{equation}
which shows the selection rule explicitly; the summation is trivial and gives a factor of 2. As before, $i$ refers to
the donor ground state, represented by an expansion of the form~\eqref{eq:expansion}. It is not surprising that the
momentum operator $\hat{p}_x$ picks up the $x$-component of the final wave vector $\kk^{(j)}$; the result
Eq.~\eqref{eq:pxmatrixelement} can of course also be obtained by acting with $\hat{p}_x$ on the plane wave
$e^{i\kk\cdot\vecti{\rho}}$ without changing to cylindrical coordinates. But in that case the selection rule will not
appear, since it is not actually present in the final result, after the summation in Eq.~\eqref{eq:pxmatrixelement} has
been carried out.

We remove the $\delta$-function in Eq.~\eqref{eq:A} by integrating with the usual 2D density of states per spin
$\mathcal{L}^2m_{||}^w/2\pi\hbar^2$. In result, we arrive at
\begin{equation} \label{eq:sigmax}
\sigma_x = \frac{2\pi^2\alpha}{\kappa\Delta k}\ \frac{\hbar K_{q=0,j}}{m_{||}^w\omega_{ji}}\ \left|
\mathcal{D}_{q=0}^{(i)}(K_{q=0,j}) \right|^2
\end{equation}
where $K_{qj}$ is defined from the energy conservation relation
\begin{equation} \label{eq:defK}
E_i + \hbar\omega_{ji} = E_{q} + \frac{\hbar^2 K_{qj}^2}{2 m_{||}^w},
\end{equation}
where $E_i$ is the energy of the donor ground state and $E_q$ the energy of the $q$-th QW level. The expansion
coefficient $\mathcal{D}_{q}^{(i)}(K_{qj})$ can be found by interpolating the matrix eigenvectors over $k$.


\section{Results} \label{sec:results}

\subsection{Numerical aspects}

In this section we present numerical results for shallow Coulombic donors in (001)-grown Si/Si$_{1-x}$Ge$_x$ quantum
wells. The electronic parameters (effective masses, band offsets and deformation potentials) of these systems are known
from the systematic study by Rieger and Vogl.\cite{rieger} We make the approximation, as discussed in
Section~\ref{sec:model}, that the difference in dielectric constant can be ignored, and use that of bulk Si throughout.
Furthermore, although the non-parabolicity of the subbands in strained Si quantum wells has been found to be
considerable, we take the basis subbands to be parabolic, since the non-parabolicity parameter is not well
known.\cite{nonpara} For small contents $x$ of Ge the effective masses are very similar for the well and barrier
regions, and therefore the discontinuity in the parallel effective mass turns out to have no observable influence on
the results for these particular systems.

For the matrix problem to be of finite order we must limit the number of QW levels to include in the basis, and also
the integration (sum) over $k$ must be cut off somewhere. This is the only real approximation in the method (other than
those inherent in effective-mass theory). Therefore, in order to assure the numerical accuracy of our results, we
always include all bound QW levels, and increase the number of unbound levels until the eigenvalues do not change. The
same procedure is applied to determine the required density and range of $k$-values. The contribution from the unbound
states drops off rapidly with increasing energy, although for narrower wells it is naturally important to include more
unbound states.

\begin{figure}[b]
\centerline{\psfig{figure=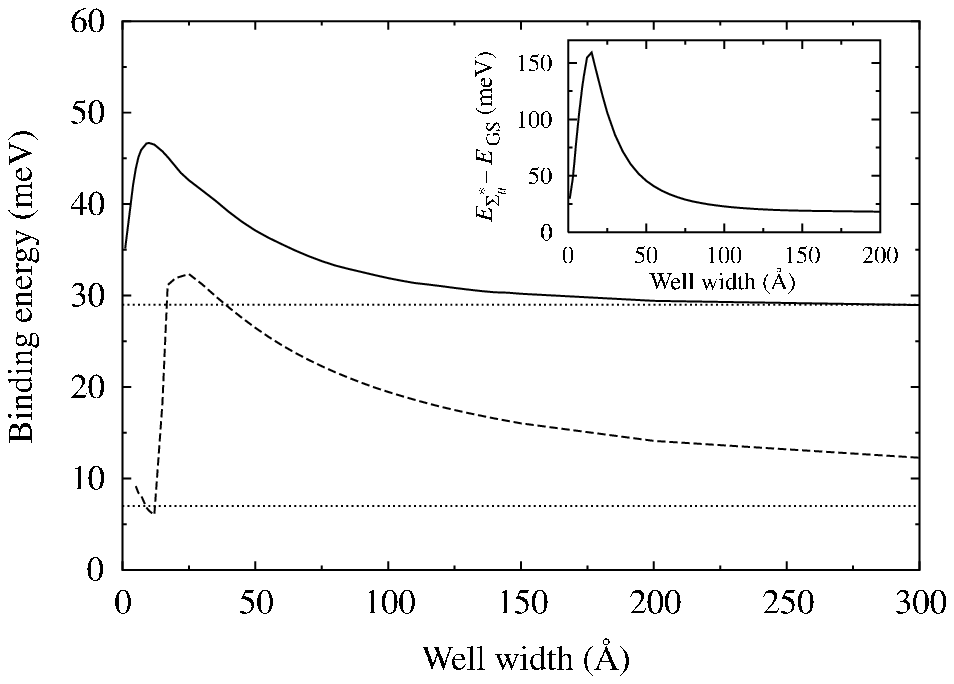,width=\linewidth}}%
\caption{The binding energy of the ground state (solid curve) and the $\Sigma_u^*$ state (dashed curve) for a
Si/Si$_{0.8}$Ge$_{0.2}$ quantum well of varying width. The impurity is located in the middle of the well, and the
central-cell shift is not included. For wide wells the binding energies approach the bulk values (29~meV and 7~meV,
respectively), indicated by the dotted lines. The inset shows the energy separation between the $\Sigma_u^*$ state and
the ground state (GS).} \label{fig:wellwidth}
\end{figure}

Quite a large range of $k$-values is needed to reach convergence, especially in the ground state wave function
amplitude, since the expansion coefficients decay rather slowly, roughly as $k^{-3.5}$. This means that, since at the
same time one wishes to keep $\Delta k$ as small as possible for accuracy, the $N\times N$ matrix problem becomes
fairly huge (typically $N$ will be between 1000 and 2000). The size of the matrix problem which can be diagonalized is
a limiting factor, and it is not always possible to obtain optimal accuracy, especially for very wide wells.
Nevertheless, we have made sure that the errors in the results presented here are less than 1\%. The only exceptions
are the energy width of the resonant state when the impurity is deep inside the barrier (in which case the width itself
is very small), and the ground state wave function amplitude; here the errors may be up to 5\%. However, the accuracy
in the {\em ratio} of the wave function amplitude to the corresponding bulk value -- which really is the quantity of
interest -- is much better than the accuracy of the amplitude itself. The wave function amplitudes are also more
sensitive to the size of the outer box $\mathcal{L}_z$ (which normalizes the "continuum" basis functions) than the
eigenvalues are. Again, the convergence can be controlled by increasing this size as much as needed.


\subsection{Binding energy and resonant state energy width}

\begin{figure}
\centerline{\psfig{figure=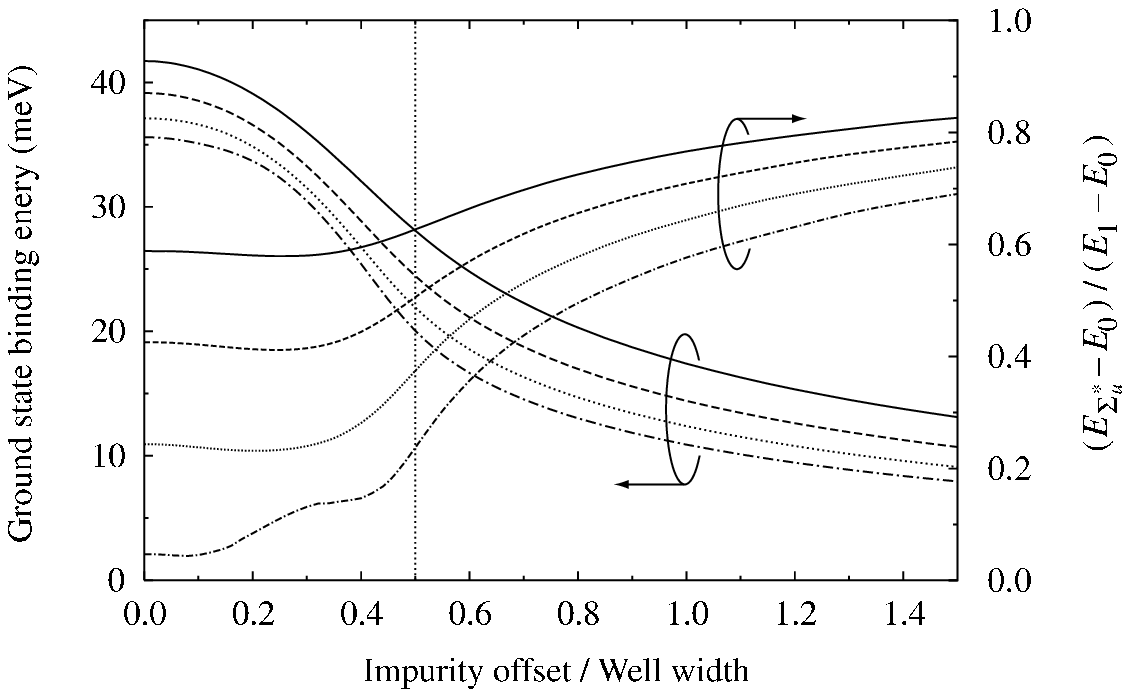,width=\linewidth}}%
\caption{The impurity ground state binding energy (left-hand vertical axis) for Si/Si$_{0.8}$Ge$_{0.2}$ wells of widths
(from top to bottom) $a$=30~\AA, 40~\AA, 50~\AA\ and 60~\AA, and for the same widths and symbols the position of the
resonant $\Sigma_u^*$ state (right-hand vertical axis). The right-hand vertical scale is normalized so that 0
corresponds to the first QW level, and 1 to the second one. The QW level energies are, from narrowest (30~\AA) to
widest (60~\AA) well: $E_0$=26.8, 17.1, 11.8, 8.7~meV and $E_1$=102.7, 67.1, 46.8, 34.4~meV. The vertical line marks
the edge of the well, and the impurity offset is measured from the middle of the well. The central-cell shift is not
included.} \label{fig:be}
\end{figure}

In agreement with variational calculations, we find that the binding energies are generally increased due to the
additional confinement presented by the quantum well potential, both for the ground and the excited states (c.f.\
Figures~\ref{fig:wellwidth} and \ref{fig:be}). Furthermore, since the $\Sigma_u^*$ state is attached to the second QW
level, it appears at higher and higher energies as the level separation increases with smaller well widths. This holds
until the well becomes very narrow, when the binding energies decrease again, as shown in Figure~\ref{fig:wellwidth}.
When the well width is much smaller than the radius of the impurity state, one can view the system as an impurity
situated in the 3D bulk, slightly perturbed by the narrow quantum well. For a vanishingly narrow well the binding
energies therefore approach the values corresponding to the bulk barrier material. Moreover, below a certain well width
there is no bound odd-parity QW level, and the $\Sigma_u^*$ state then becomes attached to the continuum, which does
not move as the well width is changed. These effects on the ground and resonant state binding energies have also been
observed in variational calculations.\cite{imagecharges,stopa}

For wide wells, the ground state binding energy converges to the effective-mass value of the bulk hydrogenic $1s$ donor
state in Si at a width of about 250~\AA, which is faster than for GaAs/Al$_x$Ga$_{1-x}$As quantum
wells.\cite{imagecharges} Clearly, the reason is the smaller radius of the impurity states in Si compared to GaAs. By
the same reasoning, the convergence of the excited $\Sigma_u^*$ state to the corresponding bulk state $2p_0$ is slower
than for the ground state, since the smaller binding energy of the excited state corresponds to a larger radius of the
state. From Figure~\ref{fig:be} we further note that for uniformly doped wells, one effectively obtains broad bands of
impurity energies, instead of a set of sharp levels. The band-width of the ground state can be almost comparable to the
spacing of the QW levels, in particular if the central-cell effect is also taken into account (see below).

The decay of the binding energies in the barrier is very slow. It follows a power-law dependence, but there does not
appear to exist any common exponent; for the curves shown in Figure~\ref{fig:be} the exponents are approximately
0.7--0.8 for the ground state and 0.2--0.4 for the resonant state. Nevertheless, at very large distances the impurity
states coalesce with the quantum well levels; see the discussion in Section~\ref{sec:classification} regarding the
impurity wave functions when the donor is placed in the barrier.

\begin{figure}
\centerline{\psfig{figure=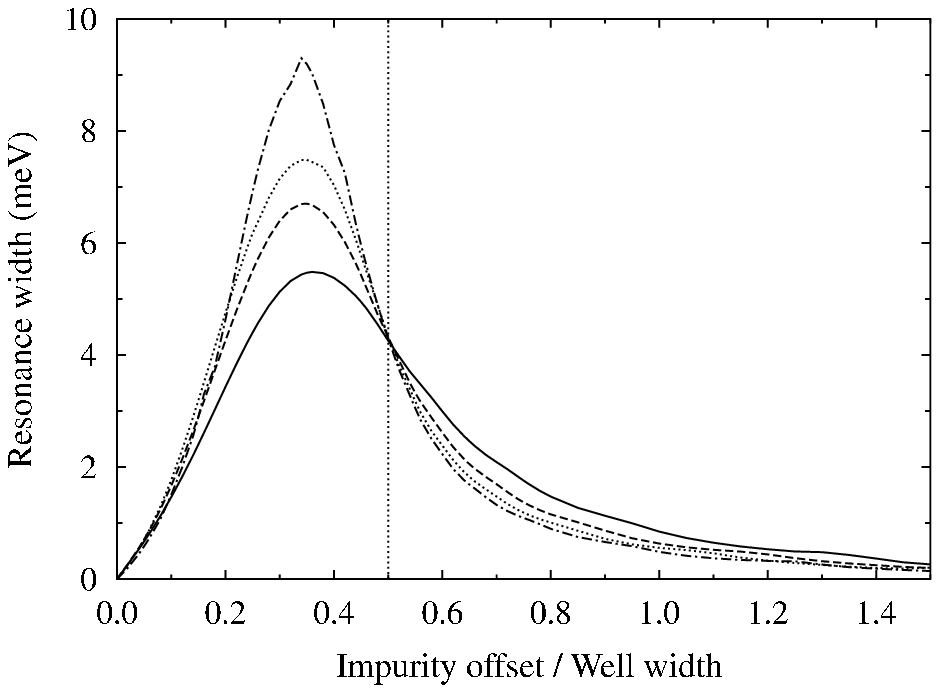,width=\linewidth}}%
\caption{The energy width $\Gamma$ of the resonant $\Sigma_u^*$ state for the same quantum wells as in
Fig.~\ref{fig:be}, with the same symbols for the curves.} \label{fig:width}
\end{figure}

Figure~\ref{fig:width} shows the energy width $\Gamma$ of the resonant $\Sigma_u^*$ state. As the impurity is moved
away from the center of the well, the width increases due to the enhanced coupling (the increased asymmetry); this
could also be seen from Figure~\ref{fig:contrib}. When we continue to move towards the barrier, $\Gamma$ reaches a
maximum value at about 35\% offset. After that it decreases, due to the reduced overlap between the two lowest QW basis
functions. In the barrier, the width continues to decay according to a power-law dependence with an exponent (which is
not common for different well widths) of the order 2--3. This is in contrast to the resonant state formed from the
shallow donor states when the impurity is in the barrier (shown in the upper-right part of Figure~\ref{fig:qw}), for
which the width decays exponentially.\cite{blomprb}

For the same relative impurity offset, the resonance is broader in wider wells, which is an effect of the fact that the
$\Sigma_u^*$ state appears closer to the bottom of the band. In Figure~\ref{fig:contrib} the same can be observed for
the higher excited $\Sigma_u$ resonant states, which are narrower the higher energy they have. We furthermore see from
Figure~\ref{fig:contrib} that a widening of the resonance causes its amplitude to decrease. The smaller the energy
width is, the more localized the impurity state is, but the effect on scattering and optical properties can still be
pronounced since the amplitude of the resonance at the same time is larger.

The behavior of the resonance position for the widest well ($a=60$~\AA) is slightly different from the other curves in
Figure~\ref{fig:be}. For this well width, the resonance is close to the subband bottom for central impurity positions.
Therefore, the resonance profile becomes rather asymmetric, since it cannot be continued below the lowest subband. The
amplitude of the resonance, as measured by $d_1(E_{\Sigma_u^*})$, is furthermore very small as long as the resonance is
close to the band bottom. At the same time the resonance width $\Gamma$ is the largest for this well width, at
intermediate impurity offsets. To accommodate this width, the resonance position shifts away from the subband bottom
more quickly in this case than for the other well widths.


\subsection{The central-cell effect}

The results discussed above for the ground state binding energy indirectly contain the effect of strain, since this
gives rise to the QW band offset, but do not include the central-cell effect, and therefore represent the
effective-mass binding energy. To take the central cell into account for the two donor states attached to the lowest
$\Delta_\bot$ valleys is however straightforward, by using Eq.~\eqref{eq:ccqwshift}. The bulk shifts $\Delta_0$,
$\Delta_E$ and $\Delta_T$ depend on the particular impurity species (values for phosphorus are given in
Section~\ref{sec:cc}), and in addition we need to know the ratio of the QW and bulk envelope functions at the impurity
position.

\begin{figure}
\centerline{\psfig{figure=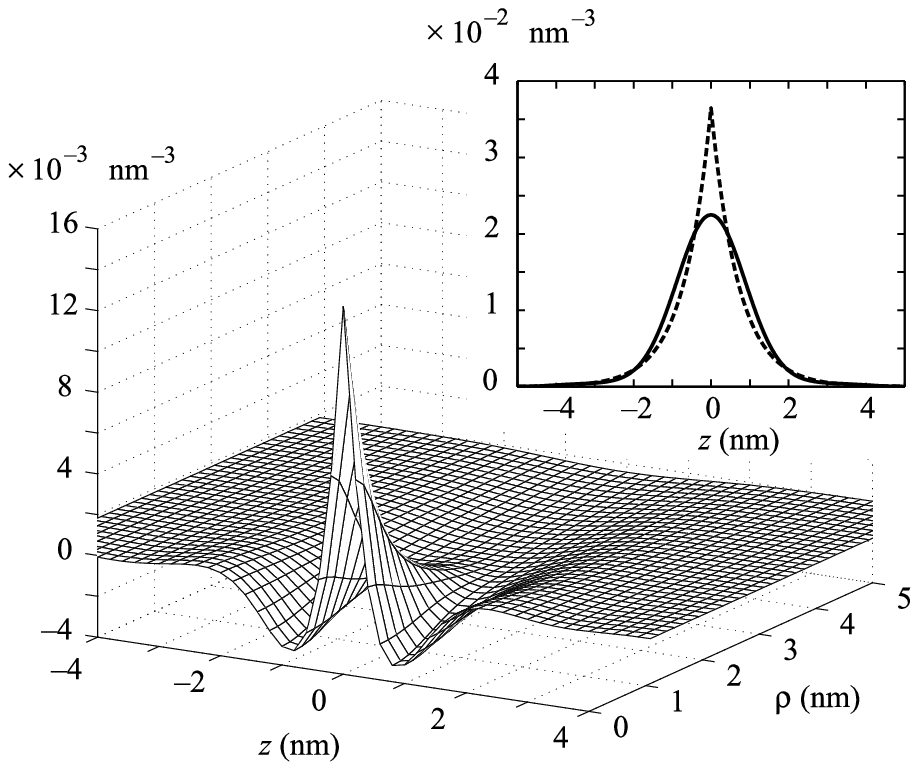,width=\linewidth}}%
\caption{The surface plot is the difference $|\varphi_{1s}|^2-|\psi_\mathrm{QW}|^2$ for a wide (200~\AA)
Si/Si$_{0.8}$Ge$_{0.2}$ quantum well, and the inset shows the QW envelope function $|\psi_\mathrm{QW}(z)|^2$ (solid)
and the bulk variational function $|\varphi_{1s}(z)|^2$ (dashed) for $\rho=0$. Both functions are normalized and give
the Si bulk binding energy (29~meV; c.f.\ Figure~\ref{fig:wellwidth}).} \label{fig:diffwavefn}
\end{figure}

Regarding the bulk envelope function, one could imagine using a variational wave function, such as the normalized
non-isotropic "hydrogenic" function
\begin{equation} \label{eq:gshydro}
\varphi_{1s}(\rho,z) = \frac{1}{\sqrt{\pi a^2b}} \exp\left( -\sqrt{\frac{\rho^2}{a^2} + \frac{z^2}{b^2}} \right),
\end{equation}
originally used by Kohn and Luttinger~\cite{KohnLuttinger} to obtain the value 29~meV for the effective-mass binding
energy in Si (the energy is minimized by $a=2.478$~nm and $b=1.420$~nm). However, although variational functions may
produce rather accurate estimates of the energy, they do not, in general, give a correct picture of the wave function.
This is illustrated by Figure~\ref{fig:diffwavefn}, where we see that in the region close to the impurity, the
variational function $\varphi_{1s}(\rho,z)$ given by Eq.~\eqref{eq:gshydro} may differ substantially from the wave
function $\psi_\mathrm{QW}(\rho,z)$ obtained with the basis expansion method, although they both reproduce the same
binding energy. The two envelope functions do however agree very well in the exponentially decaying tail, which is
precisely the part which determines the binding energy.

\begin{figure}
\centerline{\psfig{figure=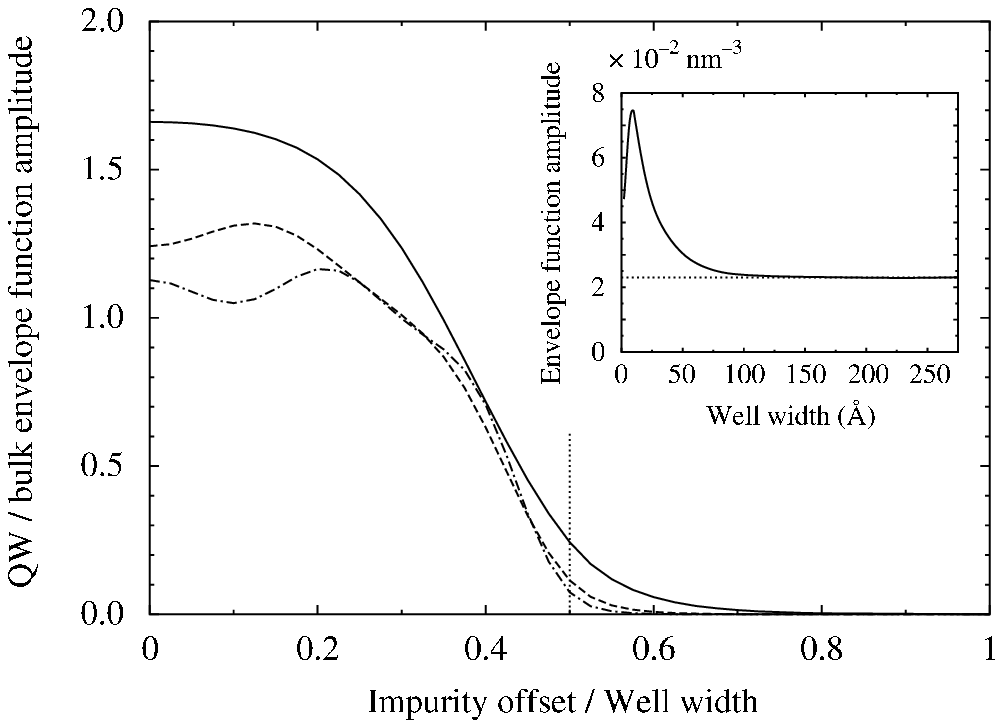,width=\linewidth}}%
\caption{The ratio of the QW envelope function amplitude to the bulk value, at the impurity position $\vect{r}_0$, for
three different Si/Si$_{0.8}$Ge$_{0.2}$ wells of widths (from top to bottom) $a$=30~\AA, 50~\AA, and 70~\AA. The
vertical line marks the edge of the well, and the impurity offset is measured from the middle of the well. The bulk
amplitude is defined as the wide-well limit (0.023~nm$^{-3}$) of the inset plot, which shows the QW envelope function
amplitude as a function of the well width. The amplitude closely follows the variation of the ground state binding
energy (c.f.\ Figure~\ref{fig:wellwidth}), also for very narrow wells when both the binding energy and the amplitude
again approach the bulk value.} \label{fig:cc}
\end{figure}

In the inset of Figure~\ref{fig:cc}, the value of the quantum well envelope function amplitude, at the impurity
position, is plotted as a function of the well width. Comparing with Figure~\ref{fig:wellwidth} we see that at a width
of 250~\AA, the ground state binding energy is extremely close to the variational value, which is reasonable since the
well is much wider than the radius of the impurity state (given by $a$ and $b$). However, the QW envelope function
amplitude differs substantially from $|\varphi_{1s}(0,0)|^2=0.0365$~nm$^{-3}$. Instead, as the well becomes wider, the
amplitude converges to a value of approximately 0.023~nm$^{-3}$, which we therefore will take as the value for the bulk
amplitude $|\phi_{1s}(\vect{r}_0)|^2$.

Once the bulk amplitude thus has been determined, we can use the ratio of the QW and bulk envelope functions, shown in
Figure~\ref{fig:cc}, to evaluate the chemical shift for any wells width and impurity position. When the impurity is
close to the center of the well, the QW wave function in narrow wells is strongly enhanced at the impurity position,
whereas for wider wells it approaches the bulk value. On the other hand, moving the impurity to the edge of the well
leads to a rapid reduction of the ratio. It was mentioned in Section~\ref{sec:classification} that when we place the
donor in the barrier, the wave function of the deep impurity states is localized in the well, and not on the impurity.
Now we find that even when the impurity is still inside the well, the wave function maximum does not exactly coincide
with the impurity position for asymmetric locations. By studying the probability density one finds that the wave
function is pushed towards the center of the well. One can view this as the wave function being reflected away by the
barriers.

For wells wider than 50~\AA, the ratio of the bulk and QW wave function amplitudes is close to unity, and hence, as was
mentioned in Section~\ref{sec:cc}, the central-cell shift is substantially smaller in the quantum well than in bulk Si.
For off-center impurity position this applies to even narrower wells (c.f.\ Figure~\ref{fig:cc}). Note also that for
very wide wells, the shift as calculated from Eq.~\eqref{eq:ccqwshift} is different from that of bulk Si, due to the
strain. Thus we can expect that the central-cell shifts are smaller in strained bulk Si than in the unstrained
material. Since the chemical shift is much larger for impurities placed in the center of the well than for positions
close to the barrier, the effective band of impurity energies mentioned above is furthermore widened by the
central-cell effect.


\subsection{Electric field and optical absorption}

\begin{figure}
\centerline{\psfig{figure=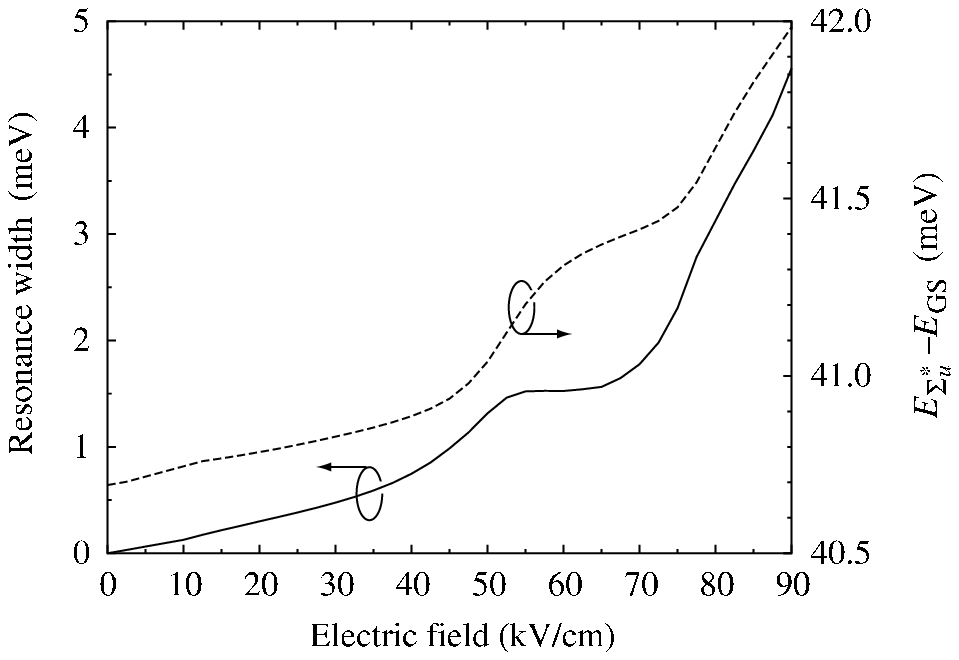,width=\linewidth}}%
\caption{The solid curve shows the energy width $\Gamma$ of the resonant $\Sigma_u^*$ state, as we apply a static
transverse electric field to a Si/Si$_{0.8}$Ge$_{0.2}$ well of width 55~\AA. The impurities are placed in the center of
the well. Also the energy separation (dashed curve) between the resonant state and the ground state (GS) is presented.}
\label{fig:efield}
\end{figure}

When a static electric field is applied across the quantum well, the parity symmetry is broken even when the impurities
are placed in the center of the well, as in the case shown in Figure~\ref{fig:efield}. The energy width $\Gamma$ of the
resonant $\Sigma_u^*$ state -- and hence the degree of coupling -- can be controlled by varying the electric field.
Also the transition energy from the resonant state to the ground state can be fine-tuned in the same way, but the
tuning range is small compared to the resonance width. By varying the QW parameters, on the other hand, this energy can
be tuned over a vast range, as shown in the inset of Figure~\ref{fig:wellwidth}.

As the electric field is increased from zero, the resonance width naturally also increases due to the induced coupling
between the resonant state and the continuum. Still, the detailed behavior of the curves in Figure~\ref{fig:efield}
warrants further comments. If we take the energy of the zero-field conduction band bottom as a fixed reference, we can
study the field-dependence of the energy positions of the impurity states and the quantum well levels (which are also
shifted by the presence of the field). The positions of the second QW level and the resonant $\Sigma_u^*$ state -- both
of which have odd parity without any applied field -- are practically unaffected by the field. On the other hand, the
lowest QW level and the impurity ground state (even parity states) are deflected downwards. Hence, as the field is
applied, the resonance appears at higher energies, relative to the subband bottom. We noted above that the energy width
is smaller for higher-energy resonances, and the interplay between this effect and the increase in the width from the
stronger coupling gives rise to the plateau at about 50--70~kV/cm. At yet higher fields, the enhancement of the
coupling will however dominate and the width increases again. Alternatively, one can consider that the electric field
"pushes" the wave functions towards the side of the well. This affects odd and even states slightly differently, which
influences the overlap between the first and second QW levels, and this is an essential factor in determining the
coupling and the resonance position.

\begin{figure}
\centerline{\psfig{figure=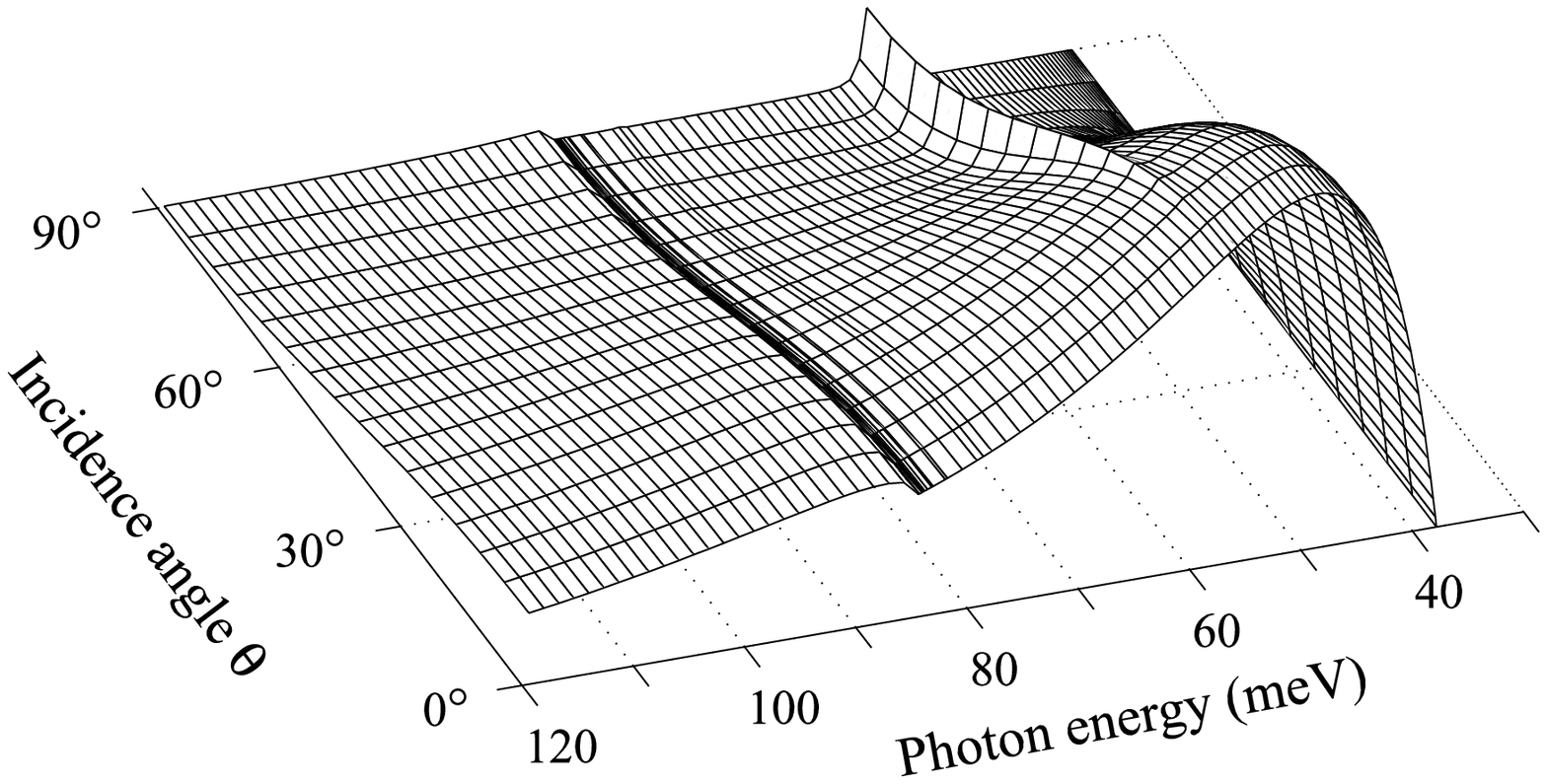,width=\linewidth}}%
\caption{The profile of the absorption cross-section $\sigma$ for a 40~\AA\ wide Si/Si$_{0.85}$Ge$_{0.15}$ quantum
well, with the impurity offset 4~\AA\ from the center of the well. The electron is initially assumed to be in the
impurity ground state; the central-cell shift has however not been taken into account for the photon energy scale. For
the actual values of the cross-section, see Figure~\ref{fig:abscut}. The coordinate system for the incidence angle
$\theta$ is defined in Section~\ref{sec:abs}.} \label{fig:abs3d}
\end{figure}

\begin{figure}
\centerline{\psfig{figure=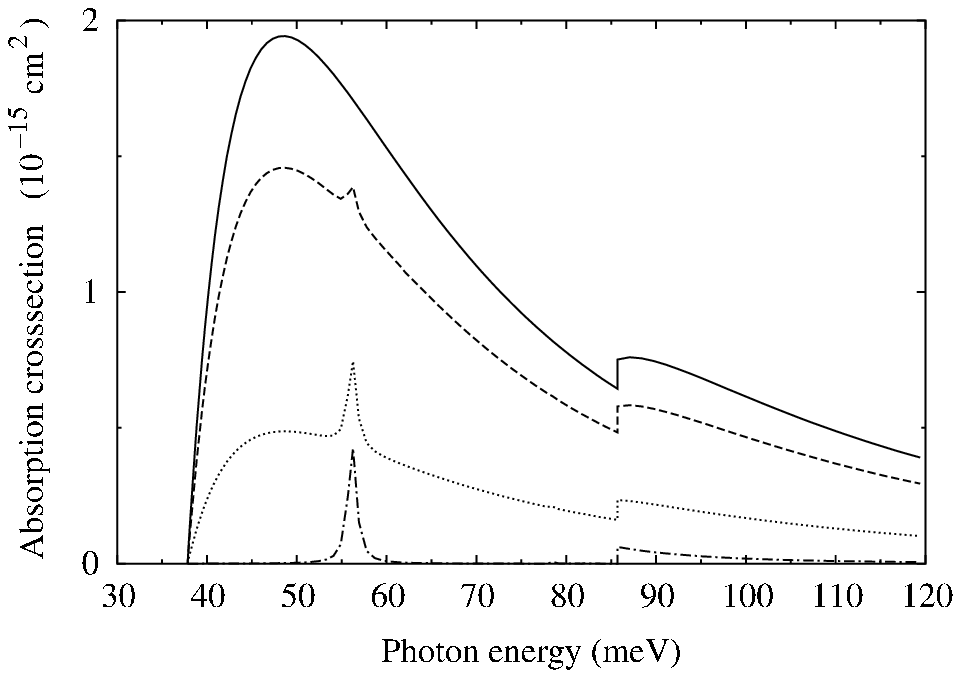,width=\linewidth}}%
\caption{For the same system as in Figure~\ref{fig:abs3d}, we here show the absorption cross-section for some selected
incidence angles $\theta$; from top to bottom $\theta$=$0^\circ$, $30^\circ$, $60^\circ$, and $90^\circ$.}
\label{fig:abscut}
\end{figure}

Finally we have also calculated the absorption cross-section $\sigma$ (c.f.\ Figures~\ref{fig:abs3d} and
\ref{fig:abscut}) from the donor ground state, in the case when the $\Sigma_u^*$ state is coupled to the first subband
by placing the impurity asymmetrically in the well. According to the dipole selection rules, discussed in
Section~\ref{sec:abs}, there is no absorption from the ground state to the unperturbed first subband ($q=0$) if the
incoming light is linearly polarized along the QW growth direction. However, when the $\Sigma_u^*$ state hybridizes
with this subband, it mixes a certain amount of the second QW subband ($q=1$) into the continuous states of the first
subband, and thus $\sigma_z\neq 0$. Hence in an energy region, determined by the width $\Gamma$, around the resonant
state, absorption is allowed even at normal incidence ($\theta=90^\circ$). This is seen in the figures as the narrow
peak at about 55~meV.

The shape of the normal incidence absorption peak is symmetric. This follows from the fact that there is no
interference between the matrix elements for absorption into the localized and continuous parts of the hybridized
resonant state, respectively (c.f.\ Section~\ref{sec:abs}). Given this, and the energy-independent 2D density of
states, the absorption spectrum will have the same shape as the resonance profile, which is symmetric in the displayed
(and nearly all other) cases, and the asymmetry often observed for resonant states~\cite{Capasso2,Fano} is hence
absent.

For angles $\theta<90^\circ$, absorption is also allowed for polarization parallel to the QW interfaces, given by
$\sigma_x$, and this gives rise to the broad background peak. Furthermore, at about 85~meV photon energy, absorption
into the second QW subband also becomes possible; in this case the selection rules allow absorption for any
polarization. The cross-section is finite at the second subband edge, reflecting the step-like 2D density of states.

The amplitude of the resonant state absorption peak is not very large compared to the background, except for
near-normal incidence ($\theta\approx 90^\circ$). Nevertheless, the absorption cross-section $\sigma_z$ is still of the
same order as the cross-section for impurity absorption in bulk Si.\cite{abscross} It is expected that the considered
intra-impurity transition is particularly strong, since the transition is analogous to the so-called resonance line
($2p\to 1s$) in atomic hydrogen.


\section{Summary and discussion} \label{sec:summary}

In this paper we have presented a unified approach for calculating the energy levels of shallow donors in
heterostructure quantum wells. By turning the Schr{\"o}dinger equation -- containing both the QW profile and the
impurity potential -- into a matrix eigenvalue problem, we obtain a complete description of the entire energy spectrum
for all donor positions, both inside and outside the well.

Applying this method to Si/Si$_{1-x}$Ge$_x$ quantum wells, we have calculated the binding energies of the ground state
and the lowest anti-symmetric resonant state for several well widths and impurity positions. The dependencies of the
binding energies on the QW parameters follow the same general behavior found in variational calculations of
GaAs/Al$_x$Ga$_{1-x}$As systems. When the donor is placed asymmetrically in the well, or a transverse electric field is
applied, the resonant state can hybridize with the continuous subbands, and we can within our method evaluate the
resonant state energy width, which is directly related to the life-time.

Si/Si$_{1-x}$Ge$_x$ quantum wells are strained due to the lattice mismatch, and this was investigated in detail, along
with the central-cell effect. By expressing the QW central-cell shift through the parameters of the split donor states
in bulk Si, and the amplitude ratio of the QW envelope function to the bulk one, it was shown that -- depending on the
impurity position and the well width -- the QW shift can be both smaller or larger compared to the bulk case.

We have also compared the commonly used variational function for donors in bulk Si, to the ground state wave function
obtained with the basis expansion. Even though both functions give the same energy if the well is wide enough, they are
far from identical. No assumption regarding the shape of the wave function is made in our non-variational method, and
we therefore conclude that the "hydrogenic" function is not capable of giving a correct description of the donor wave
function. Instead we obtain the correct envelope functions of the localized, resonant and continuum eigenstates, and
can then evaluate various matrix elements. As an example we present the ground state optical absorption spectrum, which
shows a strong dependence on the direction of the incident photon compared to the quantum well axis, due to the
selection rules.

The possibility to populate the resonant state by electrically pumping the electrons in the QW subbands, could be used
to create a conduction band resonant state laser in a Si quantum well. Provided the carriers can reach the resonance
energy without being scattered by other mechanisms, they can -- through the hybridization -- be captured into the
localized part of the resonant state. They may then make an optical transition to the impurity ground state, a
transition which is particularly strong, or to some excited localized state. Since, as was shown, the impurity states
are attached to the QW levels, it is possible to tune the intra-impurity transition energies by varying the well
parameters. The tuning range for the ground state transition was shown to extend from 25 to 150~meV by changing the Si
well width between 2 and 15~nm (the central-cell effect increases this range somewhat). Additionally, some fine-tuning
could possibly be achieved by applying a transverse electric field.

For the Si wells we have considered, the resonance width can be as large as 10~meV, which gives a very short life-time
of about 0.1~ps. One may therefore expect that the resonant states can have a strong influence not only on the
mobility, from the pronounced resonant scattering mechanism which appears in a narrow energy region, but also on the
noise spectrum, due to the capture and re-emission process. These features are in fact present even when the impurities
are placed outside the well, since there is coupling also between the shallow barrier donor state and the QW subbands.


\section{Acknowledgements}

This work was performed within the Nanometer Consortium at Lund University, and was supported by grants of the Swedish
Foundation for Strategic Research, the Swedish Research Council (TFR-THZ 2000-403), NorFA (000384), the Russian
Foundation for Basic Research, the Russian Academy of Science, and the Russian Ministry of Science. I.N.Y. is grateful
for the support from the Wenner-Gren Foundation.


\appendix

\section{Relating the central-cell parameters to the binding energies} \label{apx:cc}

In the effective-mass approximation, the wave function close to each conduction band minimum, i.e.\ in each valley
$\ell$, is assumed to be of the form $\psi_\ell(\vect{r})=\phi(\vect{r})u_\ell(\vect{r})$, with a common envelope
function $\phi$. Here $u_\ell$ is the Bloch function of the valley bottom, which can be written
$u_\ell(\vect{r})=\exp(i\kk_\ell\cdot \vect{r})U_{\kk_\ell}(\vect{r})$, where $\kk_\ell$ is the wave vector of the
respective conduction band minimum, and $U$ is a function with the periodicity of the lattice.

In bulk Si the six conduction band minima $\ell\in\{x,\overline{x},y,\overline{y},z,\overline{z}\}$ are degenerate.
Therefore we write the wave function of each $1s$ impurity state as
\begin{equation}
\Phi_{1s}^i(\vect{r}) = \phi_{1s}(\vect{r}) u_i(\vect{r}) \quad (i=1 \ldots 6),
\end{equation}
with
\begin{equation}
u_i(\vect{r}) = \sum_\ell \alpha^i_\ell\ u_\ell(\vect{r}).
\end{equation}
Using the valley bottom Bloch functions $u_\ell$ as a basis, the Bloch functions $u_i$ can be represented as vectors of
the coefficients $\alpha^i_\ell$.

However, we also know that the six $1s$ states transform as the $T_d$ group, and hence their Bloch functions $u_i$ can
be written as~\cite{BiPiBook}
\begin{equation} \label{eq:bloch1s}
\begin{split}
u_{A} &= (1,1,1,1,1,1)/\sqrt{6}, \\%
u_{E}^{(1)} &= (1,1,-1,-1,0,0)/2, \\%
u_{E}^{(2)} &= (1,1,1,1,-2,-2)/2\sqrt{3}, \\%
u_{T}^{(1)} &= (1,-1,0,0,0,0)/\sqrt{2}, \\%
u_{T}^{(2)} &= (0,0,1,-1,0,0)/\sqrt{2}, \\%
u_{T}^{(3)} &= (0,0,0,0,1,-1)/\sqrt{2}.
\end{split}
\end{equation}
The label $A$ refers to the non-degenerate ground state with energy $-\Delta_0$ relative to the effective-mass value,
$E$ is the doublet state with energy $-\Delta_0+\Delta_E$, and $T$ is the triplet state with energy
$-\Delta_0+\Delta_T$.

Since the envelope function is assumed to vary slowly over distances comparable to the range of the central-cell
potential $V$, we can write
\begin{equation} \label{eq:Voverlap1s}
\braket{\Phi_{1s}^i}{V}{\Phi_{1s}^i} = |\phi_{1s}(\vect{r}_0)|^2 \braket{u_i}{V}{u_i}
\end{equation}
where $\vect{r}_0$ is the position of the impurity. The overlap matrix elements between Bloch functions from different
valleys $\ell$ are given in Eq.~\eqref{eq:ccV}. By inserting the Bloch functions given in Eq.~\eqref{eq:bloch1s} into
Eq.~\eqref{eq:Voverlap1s}, we can express the donor energies in terms of $V_0$, $V_g$ and $V_f$ as follows:
\begin{align*}
\braket{\Psi_A}{V}{\Psi_A} &= |\phi(\vect{r}_0)|^2 \left( V_0+V_g+4V_f \right) = -\Delta_0, \\%
\braket{\Psi_E}{V}{\Psi_E} &= |\phi(\vect{r}_0)|^2 \left( V_0+V_g-2V_f \right) = -\Delta_0+\Delta_E, \\%
\braket{\Psi_T}{V}{\Psi_T} &= |\phi(\vect{r}_0)|^2 \left( V_0-V_g \right) = -\Delta_0+\Delta_T.
\end{align*}
Inverting these relationships gives the results Eq.~\eqref{eq:VfromDelta}.


\section{Donor states in strained silicon} \label{apx:largestrain}

The strain Hamiltonian for the conduction band in Si can be written~\cite{Herring}
\begin{equation}
\hat{H}_\mathrm{strain} = \Xi_d \Tr(\vect{e}) + \Xi_u (\hat{\kk}\cdot \vect{e} \cdot \hat{\kk})
\end{equation}
where $\Tr$ means the trace and $\vect{e}$ is the strain tensor. The constants $\Xi_d$ and $\Xi_u$ are the deformation
potentials, and $\hat{\kk}$ is a unit vector along one of the equivalent valleys
$\{x,\overline{x},y,\overline{y},z,\overline{z}\}$ in the unstrained material. In the basis of the valley bottom Bloch
functions (c.f.\ Appendix~\ref{apx:cc}),
\begin{multline}
\hat{H}_\mathrm{strain} =\frac{\Xi_u}{3} (e_{||}-e_\bot) \begin{pmatrix} 1&0&0&0&0&0 \\ 0&1&0&0&0&0 \\ 0&0&1&0&0&0 \\
0&0&0&1&0&0
\\ 0&0&0&0&-2&0 \\ 0&0&0&0&0&-2 \end{pmatrix} \\+\left[ \Xi_d(2e_{||}+e_\bot)
+\frac{\Xi_u}{3}(2e_{||}+e_\bot)\right] \mathbf{1}.
\end{multline}
The second term represents an overall shift and can be ignored altogether. Here $\mathbf{1}$ is the $6\times6$ unit
matrix.

For a pseudomorphically (001)-grown strained layer, the strain tensor takes the form
\begin{equation} \label{eq:straintensor}
\vect{e} = \begin{pmatrix} e_{||} & 0 & 0 \\ 0 & e_{||} & 0 \\ 0 & 0 & e_\bot
\end{pmatrix}
\end{equation}
where $e_{||}$ and $e_\bot$ are the strain tensor components parallel and perpendicular, respectively, to the interface
planes. These can be expressed in terms of the unstrained lattice constants of the layer $a_l$ and the substrate $a_s$,
by evaluating the new lattice constants in the strained layer~\cite{Walle}
\begin{equation} \label{eq:latticeconstants}
a_{||} = a_s, \qquad a_\bot = a_l \left[ 1- \frac{2C_{12}}{C_{11}} \left(\frac{a_{||}}{a_l}-1\right)\right]
\end{equation}
where $C_{ij}$ are the components of the stiffness tensor. We then have
\begin{equation} \label{eq:straintensorcomponents}
e_{||} = \frac{a_{||}}{a_l}-1, \qquad e_{\bot} = \frac{a_\bot}{a_l}-1 = - \frac{2C_{12}}{C_{11}}\ e_{||},
\end{equation}
from which we see that the two components have opposite signs.

In order to take the effects of strain and the central cell into account simultaneously, we make a unitary
transformation of $\hat{H}_\mathrm{strain}$ to the basis given by the states $u_A$, $u_E^{(i)}$ and $u_T^{(i)}$ defined
in Appendix~\ref{apx:cc}, and add the central-cell contribution
\begin{equation} \label{eq:ccshift}
\hat{H}_{\mathrm{cc}} = -\Delta_0\mathbf{1} + \begin{pmatrix} 0&0&0&0&0&0 \\ 0&\Delta_E&0&0&0&0 \\ 0&0&\Delta_E&0&0&0
\\ 0&0&0&\Delta_T&0&0
\\ 0&0&0&0&\Delta_T&0 \\0&0&0&0&0&\Delta_T \\
\end{pmatrix}
\end{equation}
which naturally is diagonal in this basis.

The resulting Hamiltonian matrix is, apart from a common diagonal constant $-\Delta_0$,
\begin{equation} \label{eq:Hccstrain}
\begin{pmatrix} 0 & 0 & -\sqrt{2}\xi & 0 & 0 & 0 \\%
0 & \Delta_E-\xi & 0 & 0 & 0 & 0 \\%
-\sqrt{2}\xi & 0 & \Delta_E+\xi & 0 & 0 & 0 \\%
0 & 0 & 0 & \Delta_T-\xi & 0 & 0 \\%
0 & 0 & 0 & 0 & \Delta_T-\xi & 0 \\%
0 & 0 & 0 & 0 & 0 & \Delta_T+2\xi
\end{pmatrix}
\end{equation}
which is easily diagonalized; the once sixfold degenerate donor ground state splits into four non-degenerate and one
twofold degenerate states (Ref.~\onlinecite{BiPiBook}, \S37), with energies
\begin{equation}
\begin{split}
\epsilon_1 &= -\Delta_0+\Delta_E-\xi, \\%
\epsilon_{2,3} &= -\Delta_0+\frac{\Delta_E}{2}\left(x+1 \pm \sqrt{1+2x+9x^2}\right), \\%
\epsilon_{4,5} &= -\Delta_0+\Delta_T-\xi, \\%
\epsilon_{6} &= -\Delta_0+\Delta_T+2\xi. %
\end{split}
\end{equation}
To abbreviate the expressions we have introduced
\begin{align}
x &= \xi/\Delta_E , \\[+1mm]%
\xi &= \Xi_u(e_\bot-e_{||})/3 = -\frac{\Xi_u}{3}\ e_{||} \left(\frac{2C_{12}}{C_{11}}+1\right). \label{eq:xi}
\end{align}
The Bloch functions of the six eigenstates are\endnote{The states $u_2$ and $u_3$ differ slightly from those given in
Eq.~(37.14) of Ref.~\onlinecite{BiPiBook}; the states given there are not, as claimed, eigenstates of the
Hamiltonian~\eqref{eq:Hccstrain}.}
\begin{equation}
\begin{split}
u_1 &= (1,1,-1,-1,0,0)/2 \\%
u_2 &= (\alpha u_{A}+u_{E}^{(2)})/\sqrt{1+\alpha^2} \\%
u_3 &= (u_{A}-\alpha u_{E}^{(2)})/\sqrt{1+\alpha^2} \\%
u_4 &= (1,-1,0,0,0,0)/\sqrt{2} \\%
u_5 &= (0,0,1,-1,0,0)/\sqrt{2} \\%
u_6 &= (0,0,0,0,1,-1)/\sqrt{2}
\end{split}
\end{equation}
expressed in the valley bottom Bloch function basis (c.f.\ Appendix~\ref{apx:cc}) with $u_{A}$ and $u_{E}^{(2)}$ given
in Eq.~\eqref{eq:bloch1s}, and
\begin{equation}
\alpha = \frac{-2x\sqrt{2}}{1+x+\sqrt{1+2x+9x^2}}.
\end{equation}

If the strain splitting $\Xi_u(e_\bot-e_{||})$ is much larger than the central-cell splitting $\Delta_E$, as is the
typical situation in Si grown on a Si$_{1-x}$Ge$_x$ substrate,\cite{rieger} $x\gg 1$ and $\alpha\approx -1/\sqrt{2}$.
In this limit the eigenfunctions $u_2$ and $u_3$ become
\begin{equation}
\begin{split}
u_2 &\approx (0,0,0,0,1,1)/\sqrt{2}, \\%
u_3 &\approx (1,1,1,1,0,0)/2,
\end{split}
\end{equation}
with energies
\begin{equation}
\begin{split}
\epsilon_{2} &\approx 2\xi -\Delta_0 +\frac{2\Delta_E}{3}, \\%
\epsilon_{3} &\approx -\xi -\Delta_0 +\frac{\Delta_E}{3}.
\end{split}
\end{equation}
Thus in this case the six states separate into two groups, where $u_2$ and $u_6$ are comprised of the two
$k_z$-valleys, and the four other states are not coupled to these valleys at all.

In the opposite limit, if the strain is very small, $x\ll 1$ and we see that $u_1$ and $u_2$ originate from the doublet
and $u_3$ from the ground state.



\clearpage


\end{document}